\documentclass[twocolumn,showpacs,floatfix,prl]{revtex4}

\usepackage{graphicx}
\usepackage{tikz}
\usepackage[caption=false]{subfig}

\usepackage{amsmath,amssymb,amsthm}
\usepackage{hyperref}

\begin{document}

\title{Topological Hopf-Chern Insulators and the Hopf Superconductor}

\author{Ricardo Kennedy\footnote[1]{ricardo.kennedy@mailbox.org}}
\affiliation{Institute for Theoretical Physics, University of Cologne, Z\"ulpicher Str. 77, 50937 Cologne}


\begin{abstract}
We introduce new three-dimensional topological phases of two-band models using the Pontryagin-Thom construction. In symmetry class $A$ these are the infinitely many Hopf-Chern topological insulators, which are hybrids of layered Chern insulators and the Hopf insulator. In symmetry class $C$, there is a $\mathbb{Z}_2$-classification with the non-trivial topological phase, the Hopf superconductor, being realized by a construction that doubles the usual Hopf insulator in momentum space. For these new topological phases we investigate the energy spectrum in the presence of a boundary and find gapless surface modes, confirming the validity of the bulk-boundary correspondence.
\end{abstract}

\maketitle

\section{Introduction}

Recently, the theoretical prediction \cite{bernevig,fukane,zhang,haldane} and subsequent experimental realization \cite{molenkamp,hasan,hasan2,chang} of various topological phases of band insulators has sparked many research efforts. On the theoretical side, the search for candidates of topological phases resulted in the ``Periodic Table for topological insulators and superconductors''~\cite{kitaev,kz}.  Given the spatial dimension as well as a set of symmetries (always including translations here) defining a symmetry class of the tenfold way \cite{hhz}, it yields the number of topological phases in the \textit{stable regime}, where the number of conduction and valence bands surpasses some lower bound. This lower bound was derived in~\cite{kz}. Below it, there are further topological phases like the topological Hopf insulators~\cite{hopf,hopf2} in three-dimensional systems of symmetry class $A$ (only translation invariance and particle number conservation) with one valence and one conduction band. 

In this paper, we expand on these findings in two directions: On the one hand, we complete the classification of topological phases in the setting of the Hopf insulators (only a subset has been studied in \cite{hopf,hopf2}) leading to what we call topological Hopf-Chern insulators; we illustrate the topological invariants of these phases through the Pontryagin-Thom construction and formulate simple tight-binding models realizing many of them. On the other hand, we explore the remaining symmetry classes: In \cite{kz} it is shown that the only other symmetry class that provides additional non-trivial strong topological phases (in the general sense introduced in \cite{charlie}) beyond the stable regime and up to three spatial dimensions is symmetry class $C$. In this symmetry class, we prove that there is a $\mathbb{Z}_2$-classification in three dimensions. We introduce a geometric diagnostic to determine this invariant based on the Pontryagin-Thom construction and provide a tight-binding model realizing the non-trivial topological phase, which we propose to call the Hopf superconductor.

\section{Hopf-Chern insulators}

For translation-invariant free fermions in symmetry class $A$, the many-body ground state of an insulator is given by assigning to each momentum the subspace of occupied states. For a two-band model with one band occupied, the occupied states are one8z-dimensional subspaces of $\mathbb{C}^2$, or equivalently points in the Grassmannian $\text{Gr}_1(\mathbb{C}^2)$. Thus, a $d$-dimensional ground state in this setting is a map
\begin{align}
T^d\to\text{Gr}_1(\mathbb{C}^2),
\end{align}
where $T^d$ is the Brillouin zone torus with coordinates $(k_1,\dots,k_d)\in[-\pi,\pi]^d$.

To group these ground states into topological phases, we impose the equivalence relation of homotopy, which is short for a continuous, or adiabatic, deformation  (see \cite{kz} for a discussion on different equivalence relations). Denoting the set of topological phases of ground state maps by $[T^d,\text{Gr}_1(\mathbb{C}^2)]$ and specializing to the case $d=3$, the set of topological phases is given by \cite{pontryagin}
\begin{align}
\begin{split}
[T^3,\text{Gr}_1(\mathbb{C}^2)]=\{(&n_0;n_1,n_2,n_3)\mid n_1,n_2,n_3\in\mathbb{Z};\\
&n_0\in\mathbb{Z}\text{ for }n_1=n_2=n_3=0\text{ and}\\
&n_0\in\mathbb{Z}_{2\cdot\gcd(n_1,n_2,n_3)}\text{ otherwise}\},\label{hopf}
\end{split}
\end{align}
where $\gcd(n_1,n_2,n_3)$ denotes the greatest common divisor of the three integers $n_1,n_2$ and $n_3$.

In \cite{hopf}, the phase $(1;0,0,0)$ is discussed and a representative is constructed that realizes it. This result is generalized in \cite{hopf2}, where representatives for all topological phases of the form $(n_0;0,0,0)$ with $n_0\in\mathbb{Z}$ are presented. Here, we complete these efforts by describing ground states in all phases of the set above, including those with non-zero invariants $n_i$ ($i=1,2,3$). The latter are obtained by restricting a representative ground state to the subtorus $T^2\subset T^3$ with $k_i=\text{constant}$ and evaluating the Chern number (= mapping degree) of the resulting map $\psi_i:T^2\to\text{Gr}_1(\mathbb{C}^2)$. For these invariants, the equivalence class containing the constant map is labeled by $n_i=0$. For the invariant $n_0$, this choice of $0$ is canonical only in the sector $n_1=n_2=n_3=0$, where $n_0$ can be reduced to the well-known Hopf invariant~\cite[pp.~227-239]{botttu}. This sector contains the constant map corresponding to the ground state in the atomic limit, the associated phase of which we call trivial. In a sector where one of $n_1,n_2,n_3$ is non-zero, the definition of the invariant $n_0$ is relative \cite{triple} in the sense that a reference ``trivial'' phase must be chosen. A natural choice from the perspective of physics (and deviating from the one chosen in \cite{triple}) is to pick the phase containing a representative realized by a layered Chern insulator. In the following part, we show that for every set of fixed values $n_1,n_2,n_3$ there indeed exists such a representative. We choose the corresponding topological phase as the required reference phase and define its invariant to be $n_0=0$. The definition of non-zero values of $n_0$ requires a generalization of the Hopf invariant. We follow \cite{triple} and use the Pontryagin-Thom construction for its definition and visualization, which we introduce presently.

\subsection{The Pontryagin-Thom construction and layered Chern insulators}

The general form of the Pontryagin-Thom construction establishes a 1-to-1 correspondence between the set $[M,S^n]$, where $M$ is an $m$-dimensional manifold, and the set of framed cobordism classes of $(m-n)$-dimensional framed submanifolds of $M$. Given a differentiable map $\psi:M\to S^n$ representing a class in $[M,S^n]$, the associated $(m-n)$-dimensional submanifold of $M$ is the preimage $\psi^{-1}(y)$ of some regular value $y\in S^n$. A framing is the pullback of some fixed basis of the tangent space at $y$ to a set of linearly independent sections of the normal bundle of $\psi^{-1}(y)$ in $M$. The framed submanifold of $M$ obtained in this way is called the \textit{Pontryagin manifold} of $\psi$. A framed cobordism between two framed $(m-n)$-dimensional manifolds $N_0$ and $N_1$ is an $(m-n+1)$-dimensional framed manifold $N\times[0,1]$ with boundaries $N\times\{0\}=N_0$ and $N\times\{1\}=N_1$. In \cite{milnor}, it is shown that two preimages of different regular values are related by a framed cobordism, that homotopies translate to framed cobordisms and finally that for all framed $(m-n)$-dimensional submanifolds of $M$ there is a corresponding map $\psi$ realizing them as a preimage. Important examples of framed cobordisms relevant to the constructions presented in this paper are illustrated in the appendix. For more examples and a more detailed introduction to this topic, see \cite{milnor}.

Since $\text{Gr}_1(\mathbb{C}^2)$ is diffeomorphic to $S^2$, we will make use of the Pontryagin-Thom construction with $M=T^d$ and $n=2$ in order to identify the set $[T^d,\text{Gr}_1(\mathbb{C}^2)]$ with framed cobordism classes of $(d-2)$-dimensional framed submanifolds of $T^d$. Starting with $d=2$, the preimage of a representative $\psi_i$ of a class in $[T^2,\text{Gr}_1(\mathbb{C}^2)]$ is a collection of points in $T^2$. A framing of a point $\mathbf{k}\in T^2$ is completely determined (up to cobordism) by the sign of $\det D\psi_i(\mathbf{k})$, where $D\psi_i(\mathbf{k})$ is the Jacobi matrix at $\mathbf{k}$. In other words, either the map is orientation-preserving (positive sign) or orientation-reversing (negative sign). This leads to the following well-known formula for the mapping degree:
\begin{align}
n_i=\sum_{\mathbf{k}\in\psi_i^{-1}(y)}\text{sgn}\det D\psi_i(\mathbf{k}),\label{eq:mapdegree}
\end{align}
where $i=1,2,3$.

\begin{figure}
\centering
\subfloat[]{\label{202const}\includegraphics[width=0.2\textwidth]{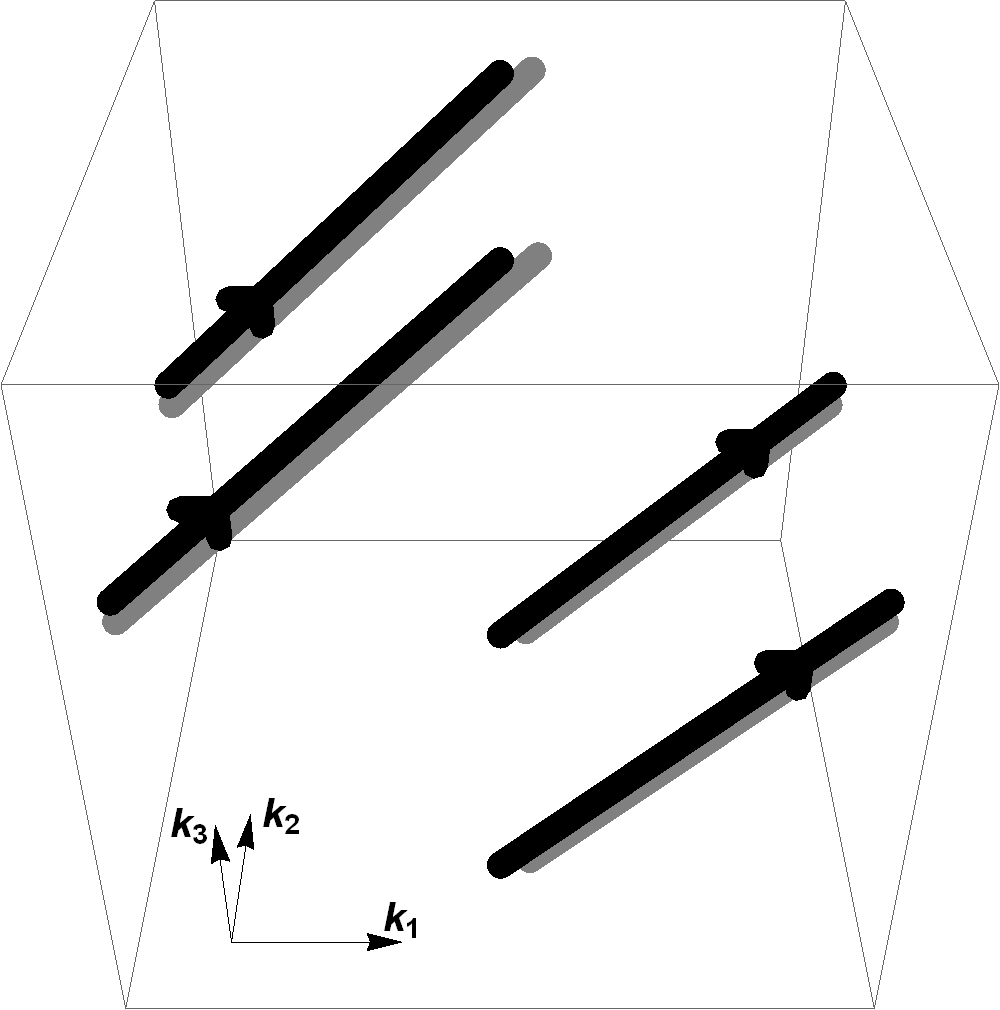}}\quad
\subfloat[]{\label{202twist}\includegraphics[width=0.2\textwidth]{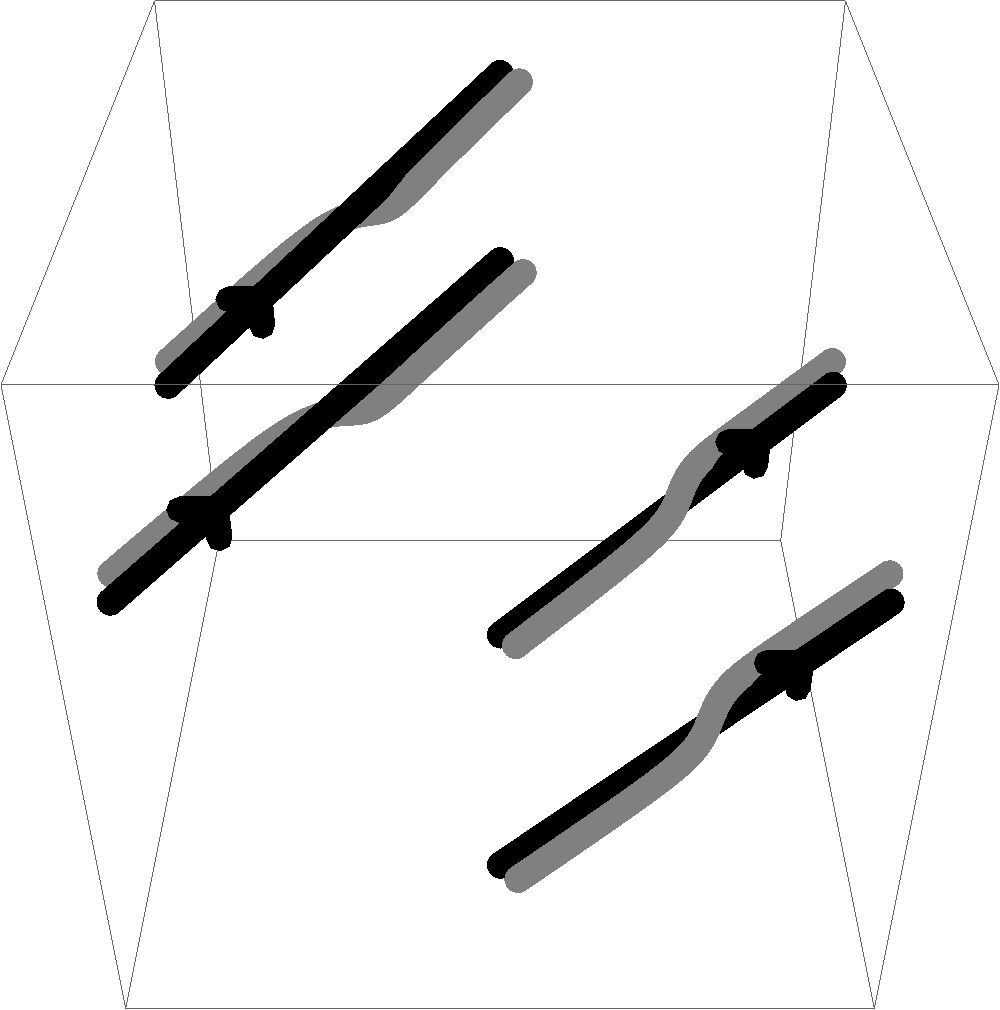}}
\caption[]{Ground state Pontryagin manifolds of two phases in the sector $\mathbf{n}=(2,0,2)$, where $n_0\in\mathbb{Z}_4$. \subref{202const} Layered Chern insulator with $n_0=0$ realized by choosing hopping directions $\mathbf{v}_1=(1,0,-1)$ and $\mathbf{v}_2=(2,2,-2)$. \subref{202twist} Corresponding Hopf-Chern insulator with Hamiltonian $\tilde H$ of eq.~\eqref{eq:Hntilde}. It has frame winding $n_0=2$, since its $\gcd(\mathbf{n})=2$ connected components have winding number 1 each.}\label{fig:202}
\end{figure}

For $d=3$, the Pontryagin manifold of $\psi$ is a one-dimensional submanifold of $T^3$. Its framing is a continuous choice of two linearly independent vectors normal to it (in other words, a pair of linearly independent sections of its normal bundle). For the purpose of visualizing such framed submanifolds, we follow \cite{triple} and equip $\psi^{-1}(y)$ with an inner orientation by requiring that the two vectors of the framing together with this inner orientation result in a fixed orientation of $T^3$ (say, right-handed). This allows us to only show one of the sections of the normal bundle (depicted by the gray line accompanying the black lines of $\psi^{-1}(y)$, see Fig.~\ref{fig:202}), since the other one can be reconstructed from the inner orientation (indicated by black arrows) in conjunction with the right-hand rule (again, up to cobordism). Thus, a ground state can now be visualized as a picture of one-dimensional black lines with arrows, which are accompanied by gray lines determining the framing (the latter may be viewed as the preimage of another regular value infinitesimally close to~$y$).

In the following, we show that every sector in the set of topological phases with fixed triplet of invariants $(n_1,n_2,n_3)\equiv\mathbf{n}$ contains a layered Chern insulator as a representative. The corresponding phase provides a natural reference phase which we define to have $n_0=0$.

The starting point for the construction of layered Chern insulators is some fixed Chern insulator with Hamiltonian $H_0(k_1,k_2)$ and associated Chern number (mapping degree) equal to 1. For instance, we may take
\begin{align}
H_0(k_1,k_2)&=\sin(k_1)\sigma_1+\sin(k_2)\sigma_2\nonumber\\
&\quad+(a+\cos(k_1)+\cos(k_2))\sigma_3\nonumber\\
&\equiv\mathbf{h}_0(\mathbf{k})\cdot\mathbf{\sigma},\label{eq:H0}
\end{align}
with some real parameter $a$ satisfying $-2<a<0$ (we choose $a=-1$ for all numerical calculations) and $\mathbf{\sigma}=(\sigma_1,\sigma_2,\sigma_3)$. This model represents a tight-binding model with only nearest neighbor hopping, but any other choice of $H_0$ with Chern number 1 can be used for the following constructions. 

As it stands, $H_0$ has no $k_3$-dependence and therefore describes a Chern insulator stacked into the $(0,0,1)$-direction. Changing this to a general direction $\mathbf{w}\in\mathbb{Z}^3$, we equip the two-dimensional layers with two new hopping directions $\mathbf{v}_1,\mathbf{v}_2\in\mathbb{Z}^3$ (replacing the original directions $(1,0,0)$ and $(0,1,0)$) which satisfy $\mathbf{v}_1\cdot\mathbf{w}=\mathbf{v}_2\cdot\mathbf{w}=0$. The new $\mathbf{k}$-dependence is then given by \cite{charlie}
 \begin{align}
H(\mathbf{k})=H_0(\mathbf{v}_1\cdot\mathbf{k},\mathbf{v}_2\cdot\mathbf{k})\label{eq:Hn},
\end{align}
where $\mathbf{k}=(k_1,k_2,k_3)\in T^3$.

If $H_0$ exhibits hopping elements of finite range on the real lattice, then so does $H$. However, the maximal hopping distance may increase in general.

We proceed by computing the triplet of invariants associated to the ground state map $\psi$ of $H$. Using the fact that $\psi$ inherits its $\mathbf{k}$-dependence from $H$, we can use eq.~\eqref{eq:mapdegree} to determine, for example, $n_3$. For this purpose, we set $k_3=0$ (or any other constant), so
\begin{align}
\psi_3(\mathbf{k})=\psi_0(A\mathbf{k}),
\end{align}
where $\psi_0$ is the ground state map of $H_0$, $\mathbf{k}=(k_1,k_2)\in T^2$ and $A_{ij}=(v_i)_j$ with $i,j=1,2$. 

Assuming that $\psi_0$ has mapping degree $1$, pick a regular value in $\text{Gr}_1(\mathbb{C}^2)$ with preimage $\mathbf{k}_0$. Then $A\mathbf{k}=\mathbf{k}_0$ has $|\det A|$ solutions for $\mathbf{k}\in T^2$ and for each of these, $\text{sgn}\det D\psi_3(\mathbf{k})=\text{sgn}\det A$. Thus, the mapping degree $n_3$ is given by
\begin{align}
n_3=\det A=(\mathbf{v}_1\times\mathbf{v}_2)_3.
\end{align}
Repeating this argument for $n_1$ and $n_2$ yields
\begin{align}
\mathbf{n}=\mathbf{v}_1\times\mathbf{v}_2.
\end{align}
Hence, all topological phases labeled by a triplet $\mathbf{n}$ can be obtained by using layered Chern insulators with suitable hopping directions $\mathbf{v}_1,\mathbf{v}_2$ (in fact, there are infinitely many possible choices). The construction above is illustrated in Fig.~\ref{fig:202}\subref{202const} for $\mathbf{n}=(2,0,2)$ .

\begin{figure}[h!]
\centering
\subfloat[]{\label{step1}\includegraphics[width=0.2\textwidth]{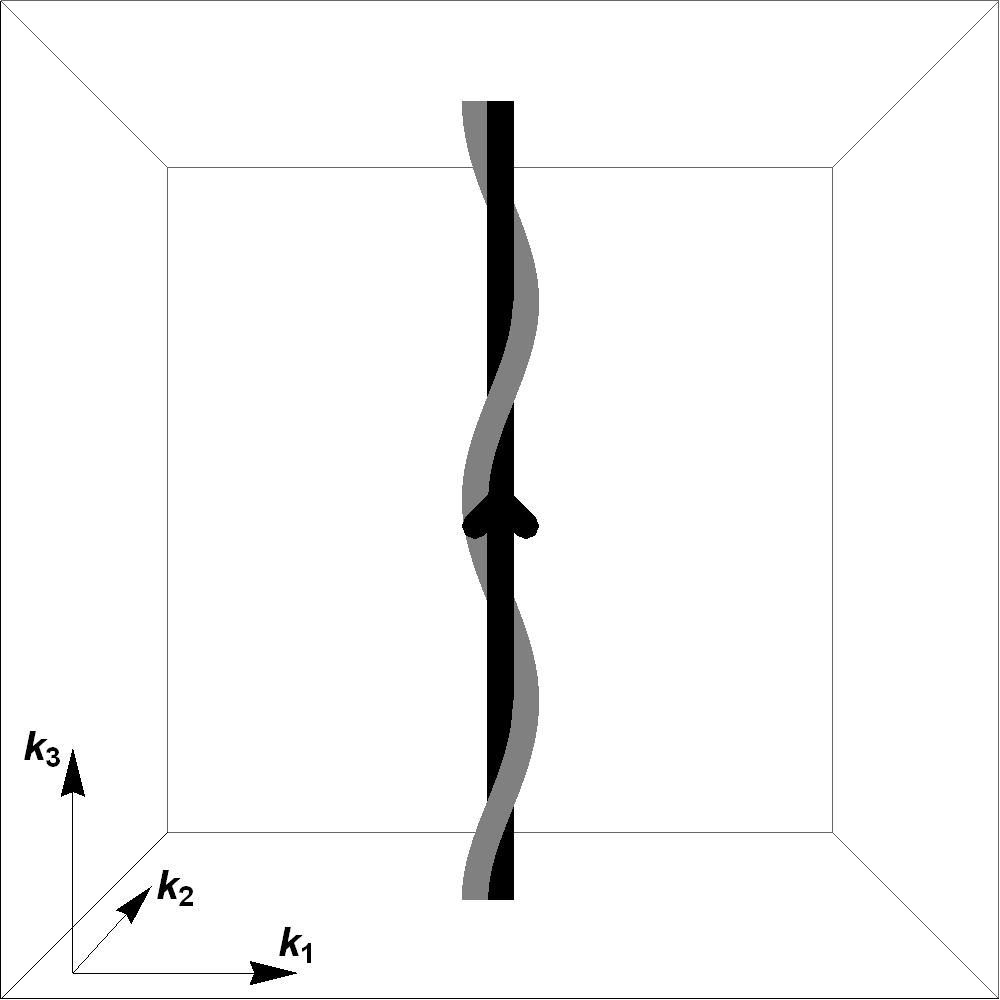}}\quad
\subfloat[]{\label{step2}\includegraphics[width=0.2\textwidth]{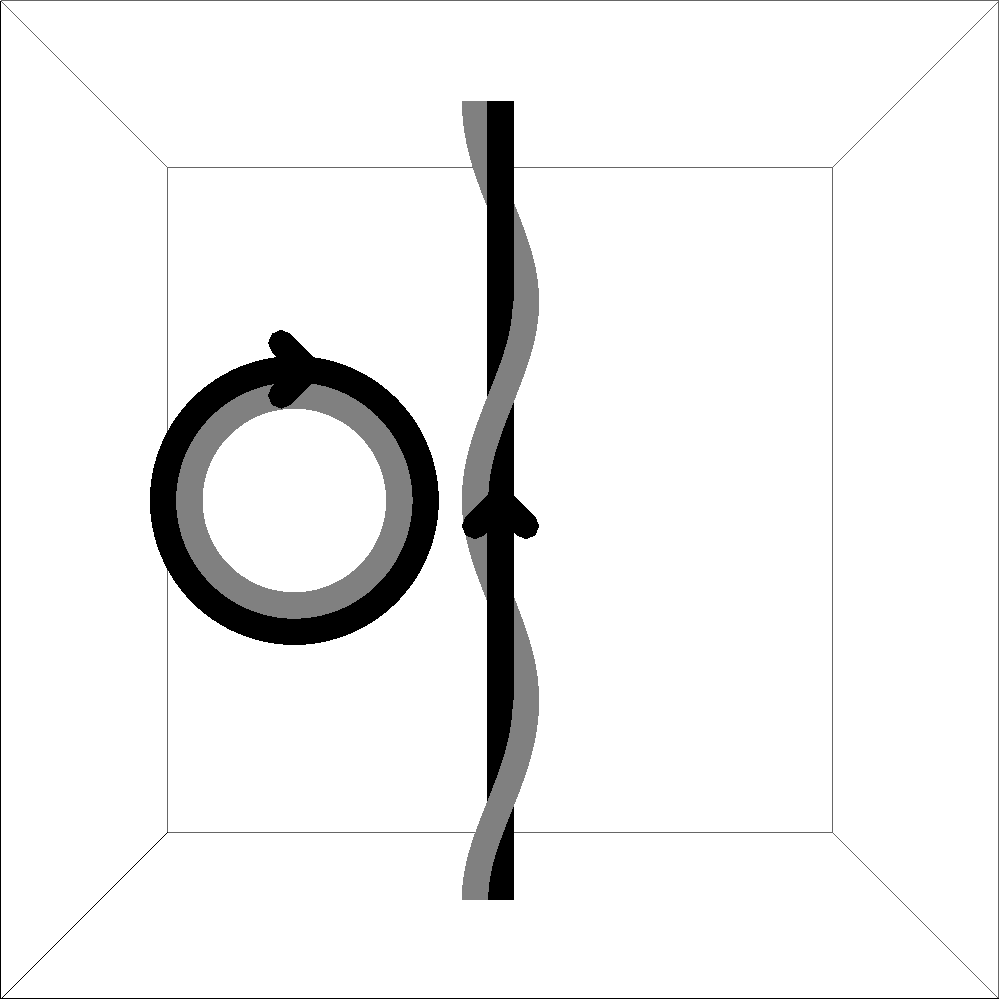}}\\
\subfloat[]{\label{step3}\includegraphics[width=0.2\textwidth]{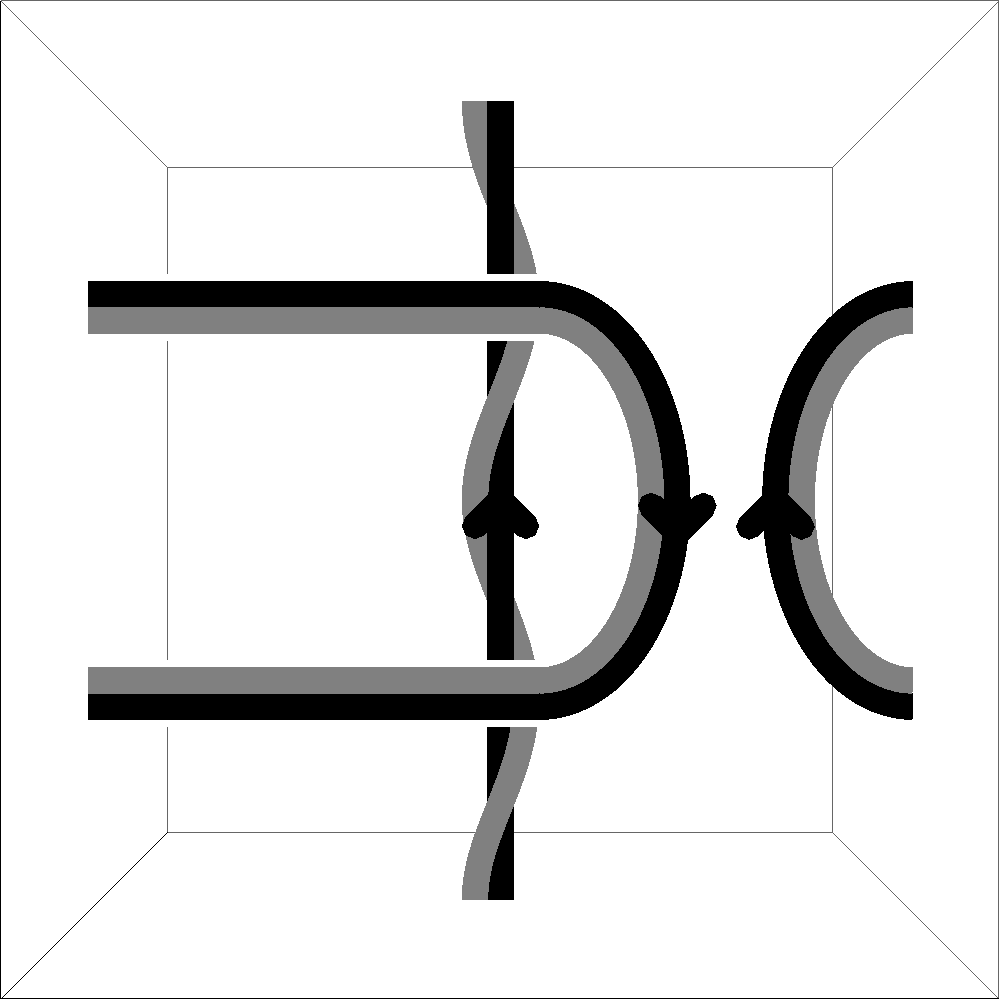}}\quad
\subfloat[]{\label{step4}\includegraphics[width=0.2\textwidth]{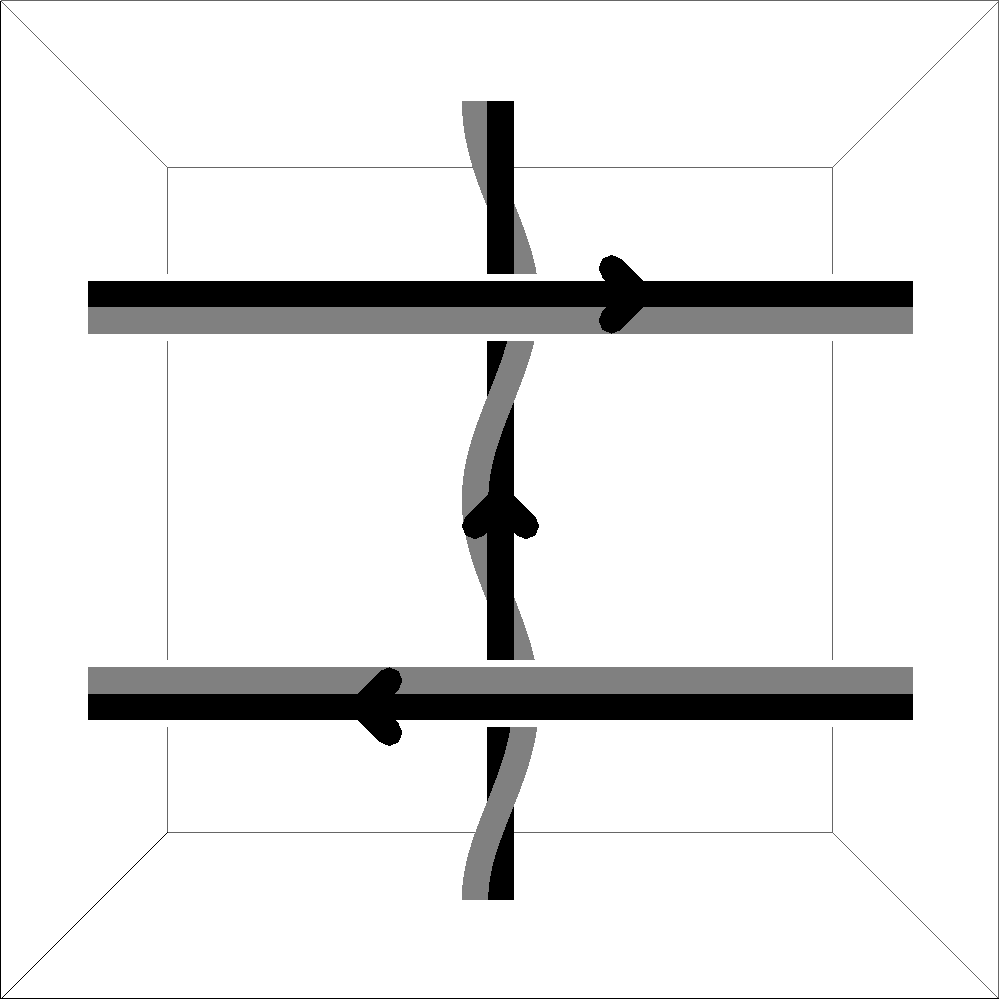}}\\
\subfloat[]{\label{step5}\includegraphics[width=0.2\textwidth]{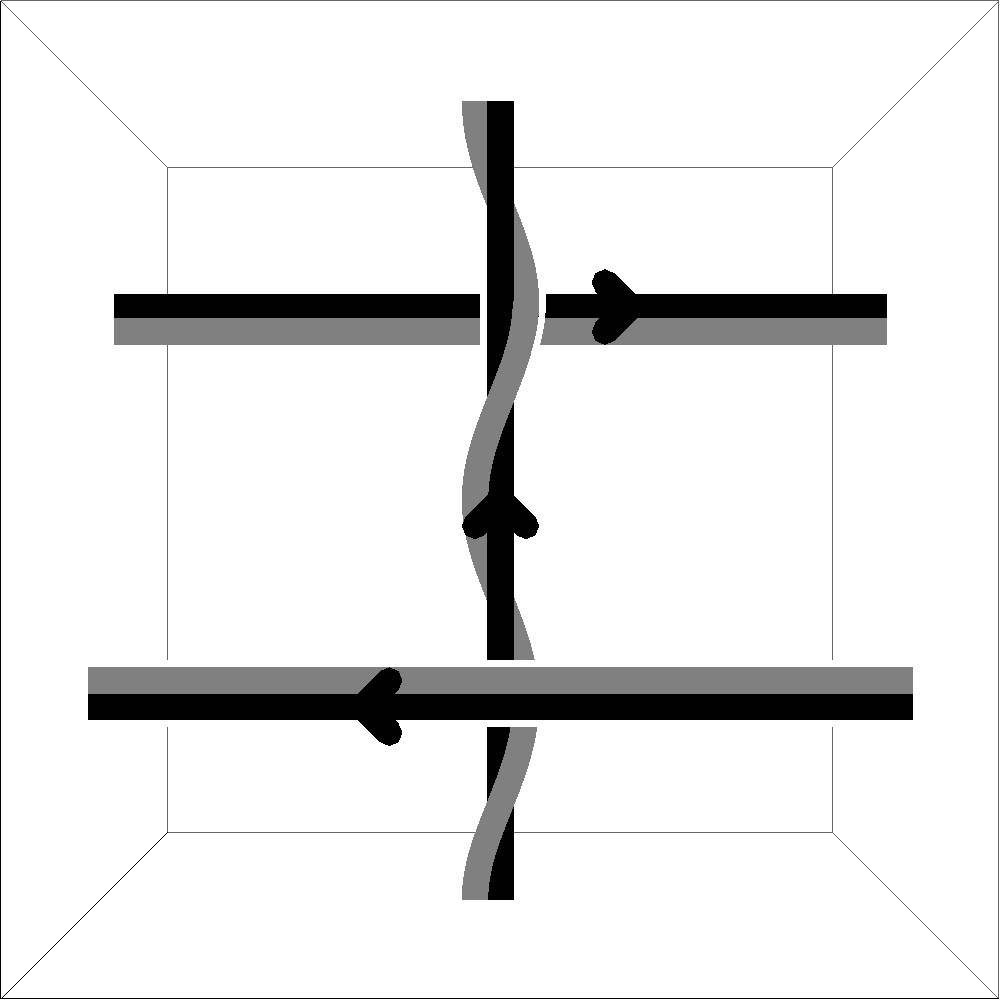}}\quad
\subfloat[]{\label{step6}\includegraphics[width=0.2\textwidth]{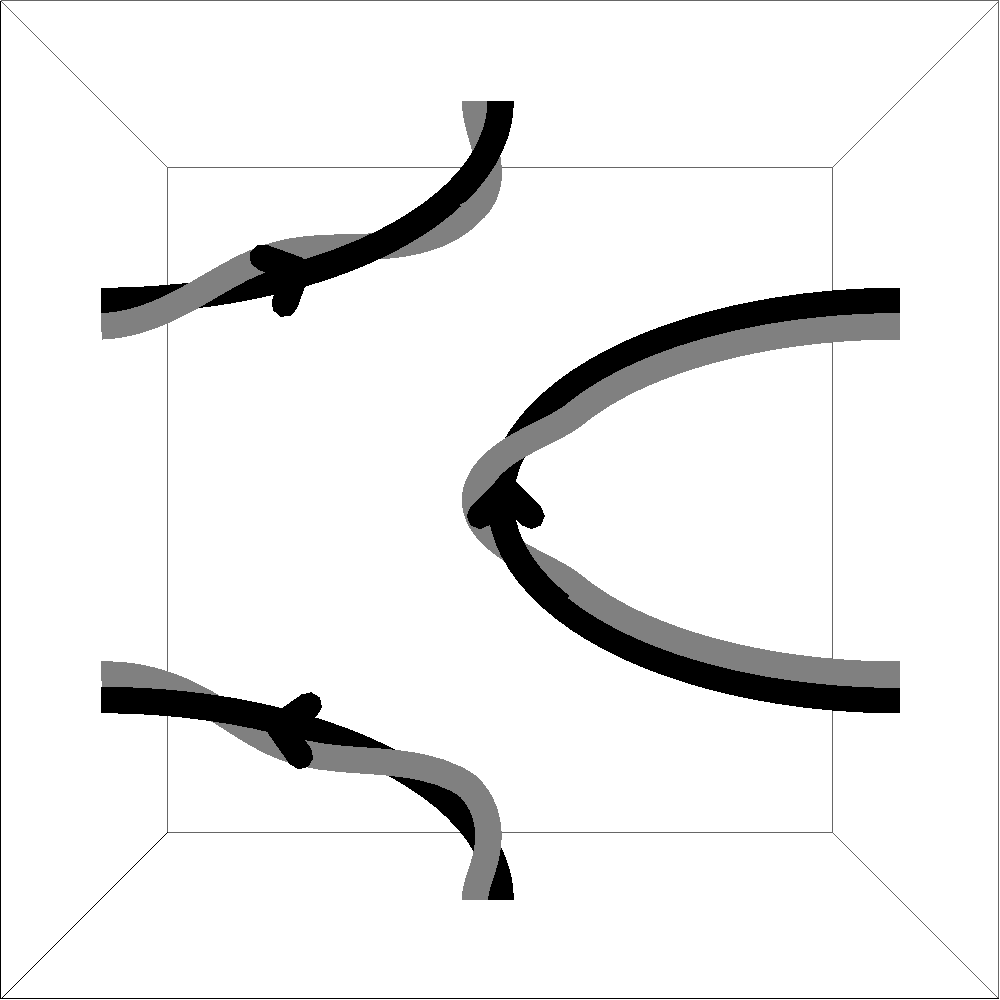}}\\
\subfloat[]{\label{step7}\includegraphics[width=0.2\textwidth]{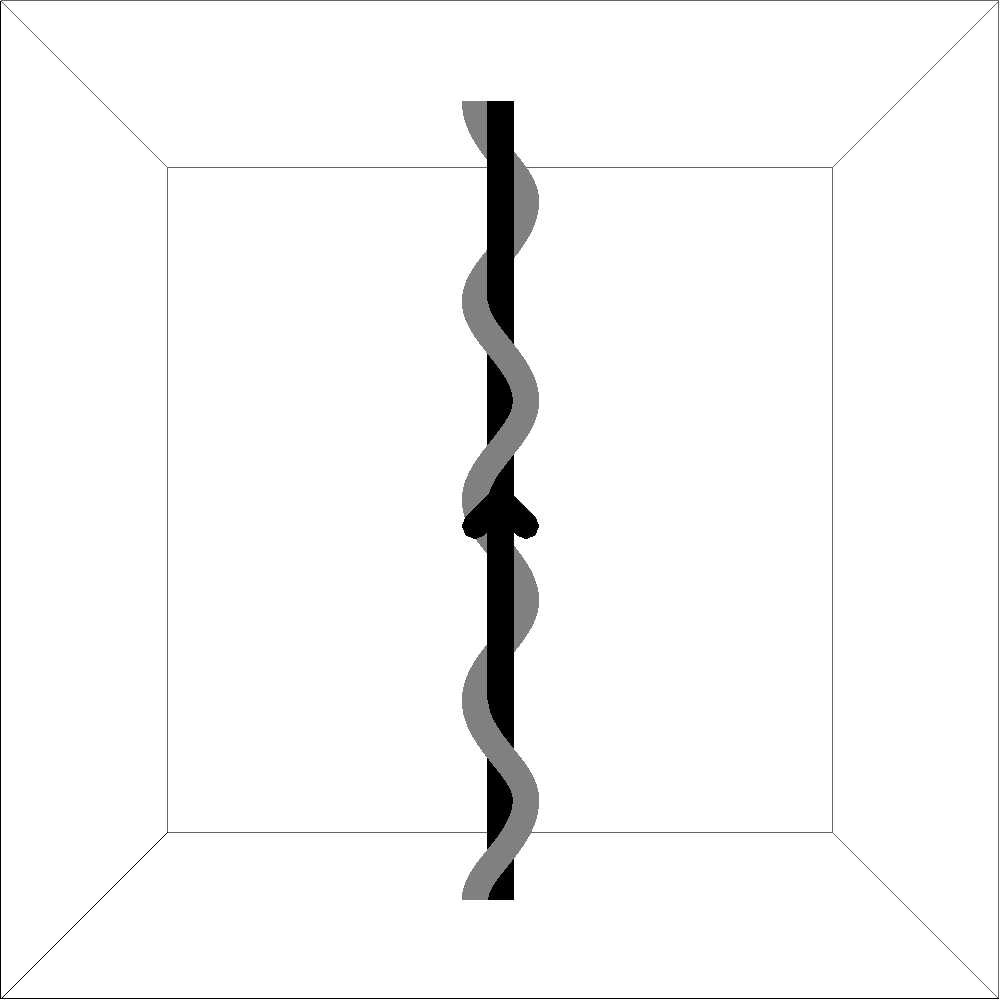}}\quad
\subfloat[]{\label{step8}\includegraphics[width=0.2\textwidth]{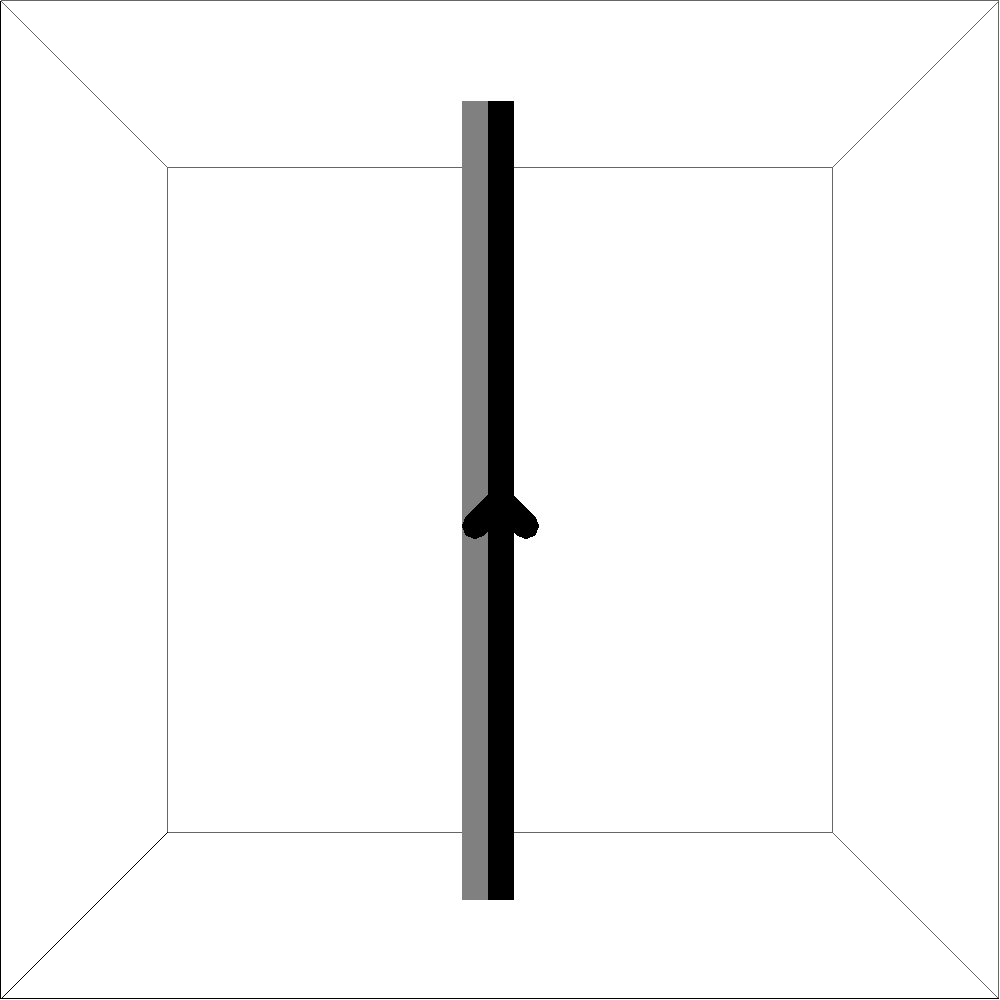}}
\caption[]{Trivialization of a frame winding $n_0=2$ for $\mathbf{n}=(0,0,1)$, demonstrating that $n_0$ is only defined modulo $2\gcd(\mathbf{n})=2$. \subref{step1}~Initial Pontryagin manifold. \subref{step2}~Cobordism Fig.~\ref{fig:cob1} used. \subref{step3}~Circle stretched using periodic boundary conditions. \subref{step4}~Cobordism Fig.~\ref{fig:cob2} used. \subref{step5}~Upper line moved using periodic boundary conditions. \subref{step6}~Cobordism Fig.~\ref{fig:cob3} used. \subref{step7}~Line ``straightened''. \subref{step8}~Framing ``straightened''. }\label{fig:double}
\end{figure}

The other $2\gcd(\mathbf{n})-1$ phases in a sector with fixed $\mathbf{n}$ and non-zero $n_0$ have no layered representatives and are in that sense ``truly three-dimensional''. For the definition of non-zero values of $n_0$ and the subsequent construction of tight-binding representatives realizing some of these topological phases, we again use the Pontryagin-Thom construction. Recall that the one-dimensional preimage of a regular value is dressed by a framing, which has been constant up to this point. In general, it is possible for the framing to wind around the preimage and $n_0$ is defined accordingly: For any ground state map one can find a homotopy such that the preimage of a regular value $y\in\text{Gr}_1(\mathbb{C}^2)$ coincides with that of the layered representative of the corresponding sector with fixed $\mathbf{n}$, up to a winding in the framing \cite{triple} (equivalently, a given preimage may be deformed using framed cobordisms to the preimage of the layered representative, up to a winding in the framing). Let the connected components of the Pontryagin manifold after this homotopy be indexed by $i$ and define $n_0^i$ to be the winding number of the framing of the $i$-th component following its inner orientation. Then the invariant $n_0$ is defined by
\begin{align}
n_0:=\sum_{i} n_0^i.
\end{align}
This winding number may be viewed as a generalization of (the infinitesimal version of) the linking number \cite{botttu} for maps $S^3\to S^2$, the famous Hopf map having a linking number of 1. It was first shown in \cite{pontryagin} (and illustrated in detail in \cite{triple}) that $n_0$ is defined only modulo $2\gcd(\mathbf{n})$ (note that in \cite{triple} the definition of $n_0$ is offset due to a different choice of reference phase with $n_0=0$). We illustrate this property for $\mathbf{n}=(0,0,1)$, where $2\gcd(\mathbf{n})=2$ and consequently there exist cobordisms deforming a frame winding $n_0=2$ to $n_0=0$, see Fig.~\ref{fig:double}.

In order to construct a tight-binding Hamiltonian for one of the $2\gcd(\mathbf{n})-1$ truly three-dimensional topological phases in a given sector $\mathbf{n}$, we again start with a reference Hamiltonian $\tilde H_0$ similar to the one introduced in eq.~\eqref{eq:H0}. Denoting by $R(\alpha)\in SO_3$ the rotation matrix around the third axis by an angle $\alpha\in[-\pi,\pi]$, we define
\begin{align}
\tilde H_0(k_1,k_2,k_3)
&:=\left(R(k_3)\mathbf{h}_0(k_1,k_2)\right)\cdot\mathbf{\sigma}\nonumber\\
&\hphantom{:}=\left(\cos(k_3)\sin(k_1)-\sin(k_3)\sin(k_2)\right)\sigma_1\nonumber\\
&\hphantom{:}\quad+\left(\sin(k_3)\sin(k_1)+\cos(k_3)\sin(k_2)\right)\sigma_2\nonumber\\
&\hphantom{:}\quad+\left(a+\cos(k_1)+\cos(k_2)\right)\sigma_3.\label{eq:H0tilde}
\end{align}
The ground state map $\tilde\psi_0$ of $\tilde H_0$ represents the topological phase $(1;0,0,1)$. We inserted $\mathbf{h}_0$ from eq.~\eqref{eq:H0} in order to give a concrete model, but any choice of $\mathbf{h}_0$ defining a ground state with Chern number (mapping degree) $1$ and regular value $y=\mathbb{C}\cdot\left(\begin{smallmatrix}0\\1\end{smallmatrix}\right)$ would work equally well. In Fig.~\ref{fig:001}\subref{hc-preimages}, the associated Pontryagin manifold $\psi^{-1}(y)$ is shown, demonstrating the winding of the framing (right) as opposed to the layered phase $(0;0,0,1)$ (left).

For an arbitrary set of invariants $\mathbf{n}=(n_1,n_2,n_3)$, we follow the prescription leading to eq.~\eqref{eq:Hn} which constructs a Chern insulator stacked into the $\mathbf{n}$-direction. However, since the starting point is now $\tilde H_0$ with a non-trivial dependence on all three momenta, we use an additional vector $\mathbf{v}_3$ parallel to $\mathbf{n}$. Since the Pontryagin manifold of the stacked representative in the sector $\mathbf{n}$ consists of $\gcd(\mathbf{n})$ connected components, we choose $\mathbf{v}_3:=\mathbf{n}/\gcd(\mathbf{n})$ in order to obtain a winding number of $1$ per component. The result of the construction is
\begin{align}
\tilde H(\mathbf{k})=\tilde H_0(\mathbf{v}_1\cdot\mathbf{k},\mathbf{v}_2\cdot\mathbf{k},\mathbf{v}_3\cdot\mathbf{k}).\label{eq:Hntilde}
\end{align}
The associated ground state map $\tilde\psi$ is a representative of the truly three-dimensional topological phase $(\gcd(\mathbf{n});\mathbf{n})$ since its Pontryagin manifold has $\gcd(\mathbf{n})$ connected components each of which is equipped with a frame winding number of $1$. In summary, we have constructed explicit models for two topological phases in each sector of invariants $\mathbf{n}$: A layered Chern insulator representing $(0;\mathbf{n})$ and a truly three-dimensional model representing $(\gcd(\mathbf{n});\mathbf{n})$ which we propose to call \textit{Hopf-Chern insulator} as it may be viewed as a hybrid of Hopf insulator and Chern insulator. The Pontryagin manifolds of representatives in the remaining $2\gcd(\mathbf{n})-2$ phases of a given sector $\mathbf{n}$ require different windings for different connected components\footnote{Cobordisms may change the number of connected components. We always refer to the representative that only differs in its framing from a layered Chern insulator.}, so the tight-binding model construction above does not apply immediately.

To illustrate our construction, we return to the example of the sector $\mathbf{n}=(2,0,2)$, where $\gcd(\mathbf{n})=2$ so that $n_0\in\mathbb{Z}_4$. Following the outlined recipe, we augment our choice of vectors $\mathbf{v}_1=(1,0,-1)$ and $\mathbf{v}_2=(2,2,-2)$ by a third vector $\mathbf{v}_3=\mathbf{n}/\gcd(\mathbf{n})=(1,0,1)$. The resulting ground state $\tilde\psi$ then represents the topological phase with $n_0=2\in\mathbb{Z}_4$. Its Pontryagin manifold is shown in Fig~\ref{fig:202}\subref{202twist}: There are $2$ connected components with a winding number of $1$ each.

\subsection{Boundary of a Hopf-Chern insulator}

\begin{figure}[h]
\centering
\subfloat[]{\label{hc-preimages}
\begin{tikzpicture}
    \def\w{0.2\textwidth}
    \def\s{4mm}
    \node[anchor=east] at (0,0)
    {\includegraphics[width=\w]{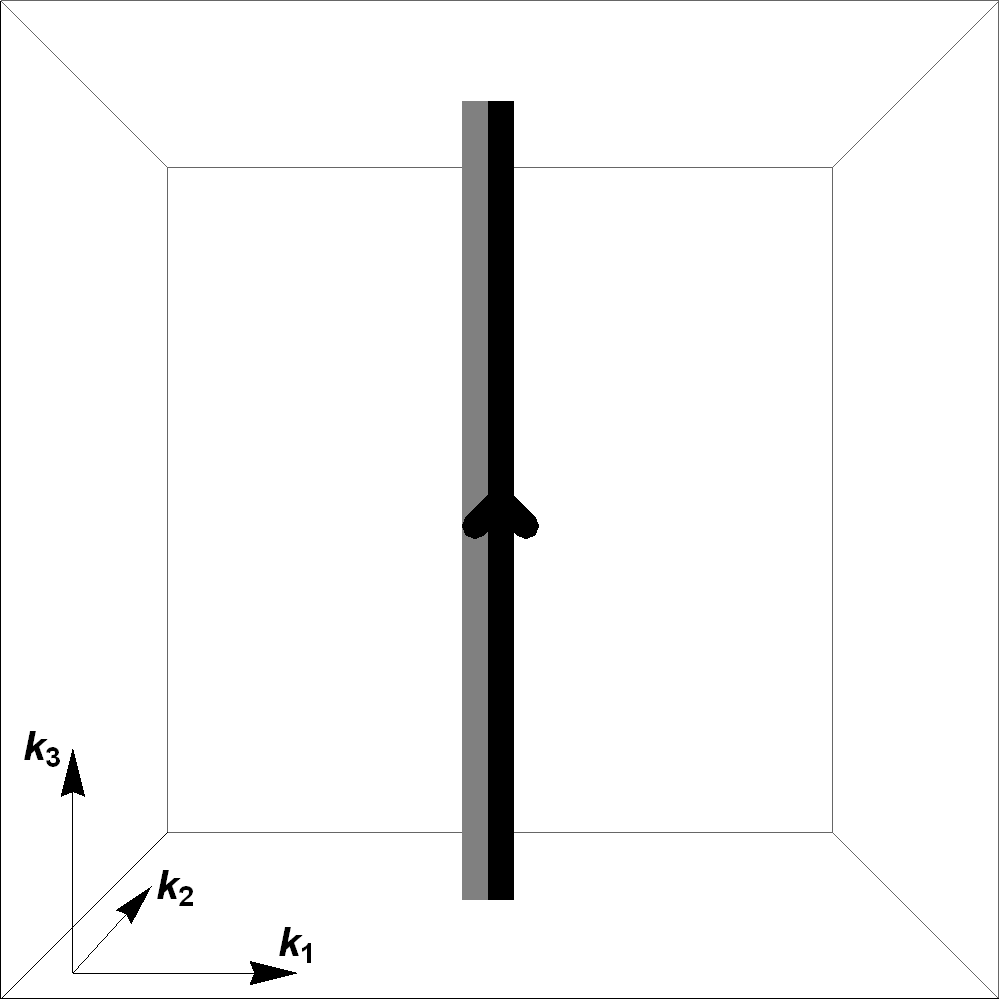}};
    \node[anchor=west] at (0,0)
    {\includegraphics[width=\w]{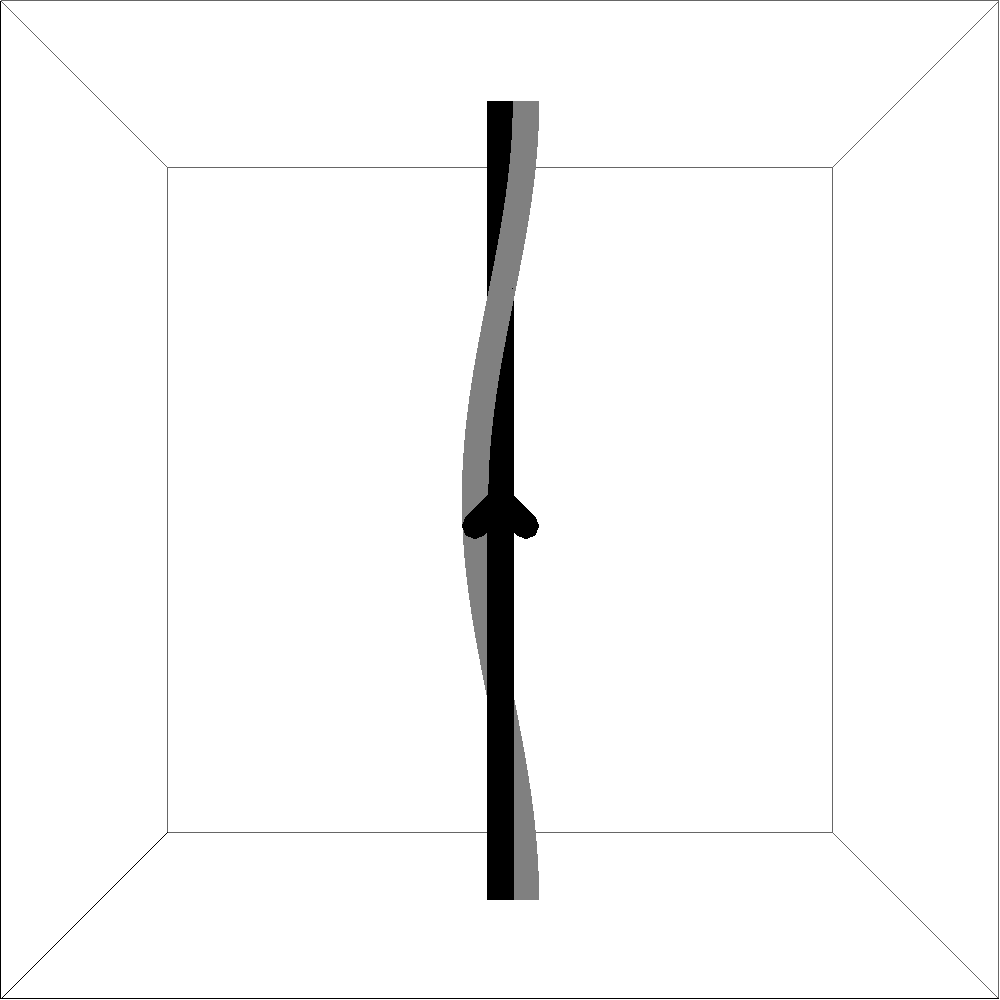}};
\end{tikzpicture}
}\\
\subfloat[]{\label{hc-x}
\begin{tikzpicture}
    \def\w{0.2\textwidth}
    \def\s{4mm}
    \node[anchor=east] at (0,0)
    {\includegraphics[width=\w]{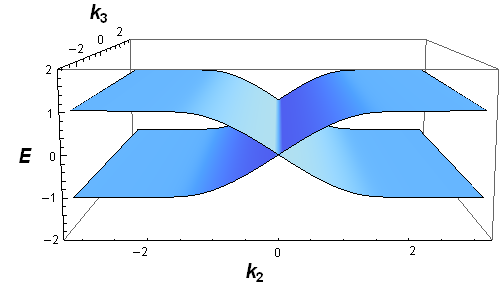}};
    \node[anchor=west] at (0,0)
    {\includegraphics[width=\w]{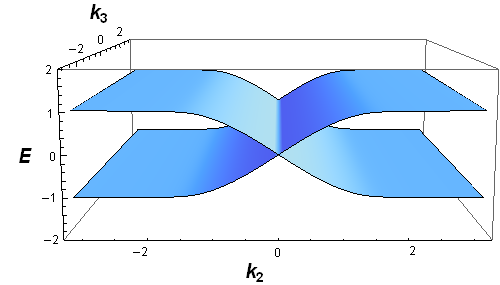}};
\end{tikzpicture}
}\\
\subfloat[]{\label{hc-z}
\begin{tikzpicture}
    \def\w{0.2\textwidth}
    \def\s{4mm}
    \node[anchor=east] at (0,0)
    {\includegraphics[width=\w]{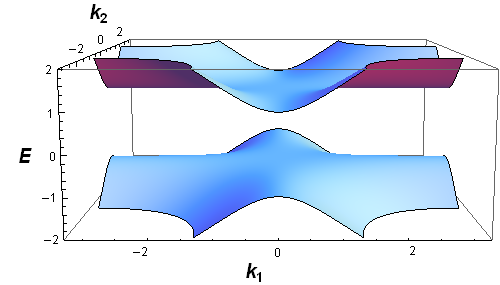}};
    \node[anchor=west] at (0,0)
    {\includegraphics[width=\w]{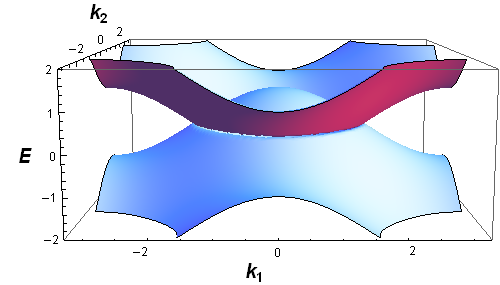}};
\end{tikzpicture}
}
\caption[]{Comparison of phases $(0;0,0,1)$ (left column) and $(1;0,0,1)$ (right column) realized by the Hamiltonians in eq.~\eqref{eq:H0} resp. eq.~\eqref{eq:H0tilde}.\subref{hc-preimages} Pontryagin manifolds of regular value $\mathbb{C}\cdot\left(\begin{smallmatrix}0\\1\end{smallmatrix}\right)$. \subref{hc-x}~Energy spectra with boundary orthogonal to $(1,0,0)$ for a slab with $16$ sites. Shown are the two energy eigenvalues closest to the Fermi energy $E=0$. For a boundary orthogonal to $(0,1,0)$ the spectra look identical with $k_2$ replaced by $k_1$. \subref{hc-z}~Energy spectra with boundary orthogonal to $(0,0,1)$: Gapped for the layered Chern insulator (left) and gapless with a ring of zero modes for the Hopf-Chern insulator (right).}\label{fig:001}
\end{figure}

For the purpose of numerically investigating the bulk-boundary correspondence, we return to the simplest example given by the topological phase $(1;0,0,1)$. A Hamiltonian realizing this phase was introduced in eq.~\eqref{eq:H0tilde} (note that the construction described above also applies in this case with $(\mathbf{v}_i)_j=\delta_{ij}$). We introduce boundaries by using partial inverse Fourier transformations (see~\cite{tft}) and subsequently truncating the hopping Hamiltonian. This effectively puts the system on a slab geometry with two (infinitely extended) surfaces where two independent momentum coordinates remain good quantum numbers. In particular, we investigate boundaries perpendicular to the three directions $(1,0,0)$, $(0,1,0)$ and $(0,0,1)$ and compare the results with the layered Chern insulator of the $(0;0,0,1)$-phase, see Fig.~\ref{fig:001}.

For the latter, boundaries in the $(1,0,0)$ and $(0,1,0)$ produce Dirac ``valleys'' in the spectrum and in particular the energy gap closes. This comes as no surprise, since the layered Chern insulator has no $k_3$-dependence and its boundary spectrum is simply the one obtained in a two-dimensional strip geometry copied along the $(0,0,1)$-direction. The spectrum for the Hopf-Chern insulator in the phase $(1;0,0,1)$ with Hamiltonian $\tilde H_0$ is similar in these cases, see Fig.~\ref{fig:001}\subref{hc-x}.

However, a sharp distinction emerges with a boundary orthogonal to the $(0,0,1)$-direction. For the layered Chern insulator, the spectrum is simply that of a two-dimensional Chern insulator with a degeneracy equaling the number of layers. Thus, by definition, the energy gap does not close, since otherwise the uncoupled layers would not be insulators in the first place. In contrast, the spectrum of the Hopf-Chern insulator shows a ring of zero modes similar to the one shown in~\cite{hopf} for the Hopf insulator, demonstrating its inherent three-dimensional nature and the apparent applicability of bulk-boundary correspondence.

\section{The Hopf superconductor}

The Hopf-Chern insulators discussed in the previous part reside in symmetry class $A$. In this section, we turn our attention to symmetry class $C$ in three dimensions, realized for instance by spin-singlet superconductors. We consider the case of a superconductor formed from a single band with annihilation operators $\gamma_\mathbf{k}=u(\mathbf{k})c^\dagger_{-\mathbf{k},\downarrow}+v(\mathbf{k})c_{\mathbf{k},\uparrow}$. Thus, the subspace of annihilation operators defining a ground state \cite{kz} is $1$-dimensional in this case, leading again to the target space $\text{Gr}_1(\mathbb{C}^2)$. However, the canonical anti-commutation relations impose a relation between annihilation operators at $\pm\mathbf{k}$, so all ground state maps need to be $\mathbb{Z}_2$-equivariant \cite{kz}. The set of topological phases in this case is denoted by $[T^3,\text{Gr}_1(\mathbb{C}^2)]^{\mathbb{Z}_2}$. On $T^3$, the $\mathbb{Z}_2$-action is generated by $\mathbf{k}\mapsto-\mathbf{k}$ and on the Grassmannian $\text{Gr}_1(\mathbb{C}^2)$ it happens to be trivial. The latter is a peculiarity of the restriction to a single band and in general the $\mathbb{Z}_2$-action is non-trivial with its generator fixing the points of a subset~$\text{Sp}_{2n}/\text{U}_n\subset\text{Gr}_n(\mathbb{C}^{2n})$. 

 In \cite{charlie} we showed that  the set $[S^d,\text{Gr}_1(\mathbb{C}^2)]^{\mathbb{Z}_2}$ obtained by replacing the Brillouin zone torus $T^d$ by the $d$-dimensional sphere $S^d$ is in bijection with the subset of $[T^d,\text{Gr}_1(\mathbb{C}^2)]^{\mathbb{Z}_2}$ that is trivial on any $\mathbb{Z}_2$-invariant subtorus $T^{d-1}$. This bijection can be established by modeling $T^d$ as a cube $I^d=[-\pi,\pi]^d$ with periodic boundary conditions and $S^d$ as the quotient $I^d/\partial I^d$. A class that is trivial on all $\mathbb{Z}_2$-invariant subtori $T^{d-1}$ has a representative that is constant on the boundary $\partial I^d$, so it may be viewed as a map from $I^d/\partial I^d=S^d$. Importantly, the bijection holds for all band numbers, so in particular beyond the stable regime. In symmetry class $A$, which can be included in the equivariant setting with trivial $\mathbb{Z}_2$-actions, this bijection is between the subset with invariants $(n_0;0,0,0)$ investigated in \cite{hopf} and \cite{hopf2} (in these works, the triviality on subtori $T^2\subset T^3$ is checked explicitly) and $[S^3,\text{Gr}_1(\mathbb{C}^2)]=\pi_3(\text{Gr}_1(\mathbb{C}^2))=\mathbb{Z}$. 
 
With $T^d$ replaced by $S^d$ and $d\le3$, it is shown in \cite{kz} that among all symmetry classes -- besides the Hopf insulators in class $A$ -- only symmetry class $C$ provides an additional topological phase beyond the stable regime. In fact,
\begin{align}
[S^3,\text{Gr}_n(\mathbb{C}^{2n})]^{\mathbb{Z}_2}=
\begin{cases}
\mathbb{Z}_2&\text{for $n=1$}\\
0&\text{for $n>1$.}
\end{cases}
\end{align}
For $n>1$ the stable classification holds \cite{kz} and the corresponding entry in the Periodic Table shows that there is only the trivial topological phase. For $n=1$, a separate treatment is required, for which we will make use of some concepts from homotopy theory (for an introduction, see e.g. \cite{hatcher}). Since the $\mathbb{Z}_2$-action for $n=1$ is trivial on the target space $\text{Gr}_1(\mathbb{C}^{2})=S^2$, a ground state map $\psi$ satisfies $\psi(\mathbf{k})=\psi(-\mathbf{k})$. We may discard this equivariance condition by replacing $S^3$ with the quotient $S^3/\mathbb{Z}_2=\Sigma S^2/\mathbb{Z}_2=\Sigma\mathbb{R}P^2$, where $\Sigma$ denotes the suspension:
\begin{align}
[S^3,\text{Gr}_1(\mathbb{C}^{2})]^{\mathbb{Z}_2}
&=[\Sigma\mathbb{R}P^2,\text{Gr}_1(\mathbb{C}^{2})]\nonumber\\
&=[\Sigma\mathbb{R}P^2,S^2].
\end{align}
Since the map $f:S^1\to S^1$, $f(\phi)=2\phi$ (where $\phi$ is the angle coordinate) has a mapping cone $C(f)=\mathbb{R}P^2$, there is an associated long exact Puppe-sequence \cite[p.~398]{hatcher} containing the following part:
\begin{align}
[\Sigma^2 S^1,S^2]&\to[\Sigma^2 S^1,S^2]\to[\Sigma\mathbb{R}P^2,S^2]\nonumber\\
&\to[\Sigma S^1,S^2]\to[\Sigma S^1,S^2].
\end{align}
The first map is $(\Sigma^2f)^*$, followed by $(\Sigma p)^*$, where $p:C(f)\to\Sigma S^1$ collapses $S^1\subset C(f)$ to a point. The third map is $(\Sigma i)^*$ induced by the inclusion $i:S^1\hookrightarrow C(f)$ and the last map is $(\Sigma f)^*$. Using $\Sigma^nS^1=S^{n+1}$ the sequence simplifies to
\begin{align}
\pi_3(S^2)\to\pi_3(S^2)\to[\Sigma\mathbb{R}P^2,S^2]\to\pi_2(S^2)\to\pi_2(S^2).
\end{align}
The homotopy group $\pi_3(S^2)$ is isomorphic to $\mathbb{Z}$ and classified by the Hopf invariant. Similarly, $\pi_2(S^2)\simeq\mathbb{Z}$ is classified by the mapping degree. Since $f$ doubles the angle of $S^1$, the induced maps $(\Sigma^2 f)^*$ and $(\Sigma f)^*$ are simply multiplication by $2$ on $\mathbb{Z}$. Exactness then leads us to the result
\begin{align}
[S^3,\text{Gr}_1(\mathbb{C}^{2})]^{\mathbb{Z}_2}=[\Sigma\mathbb{R}P^2,S^2]=\mathbb{Z}_2.
\end{align}
The derivation presented here immediately delivers representative ground states realizing the non-trivial phase. A representative of a class in $\pi_3(S^2)\simeq\mathbb{Z}$ with odd Hopf invariant is mapped to a representative of the non-trivial element of $\mathbb{Z}_2$ via $(\Sigma p)^*$. The result is a map that is constant on the subset $\Sigma S^1\subset \Sigma C(f)=\Sigma\mathbb{R}P^2$. 

From this representative, we now retrace the steps of removing the equivariance condition as well as replacing $T^3$ by $S^3$ in order to obtain an equivariant map $\psi:T^3\to\text{Gr}_1(\mathbb{C}^{2})$. The sphere $S^3=\Sigma^2S^1$ corresponds to $\tilde I^3/\partial\tilde I^3$, where $\tilde I^3$ is half of $I^3$, defined by e.g. $k_1\le0$. We first lift the map with odd Hopf invariant to a map with domain $\tilde I^3$, which is constant on $\partial\tilde I^3$. Then, we reinstate the equivariance condition by extending to the entire cube $I^3$ according to the rule $\psi(\mathbf{k})=\psi(-\mathbf{k})$. Finally, we interpret the resulting map as a map from $T^3$. Importantly, in the step reinstating the equivariance condition the odd Hopf invariant we started with is reversed on the other half of the domain, so the total Hopf invariant is $0$. Because of the prominent role of the Hopf invariant in the construction of this non-trivial topological superconductor, we call this phase the \textit{Hopf superconductor}. Its construction resembles that of time-reversal invariant topological insulators in two-dimensional systems of symmetry class $A$II, where non-trivial representatives may also be created by doubling representatives with odd Chern numbers such that the total Chern number vanishes.

\begin{figure}
\centering
\includegraphics[width=0.4\textwidth]{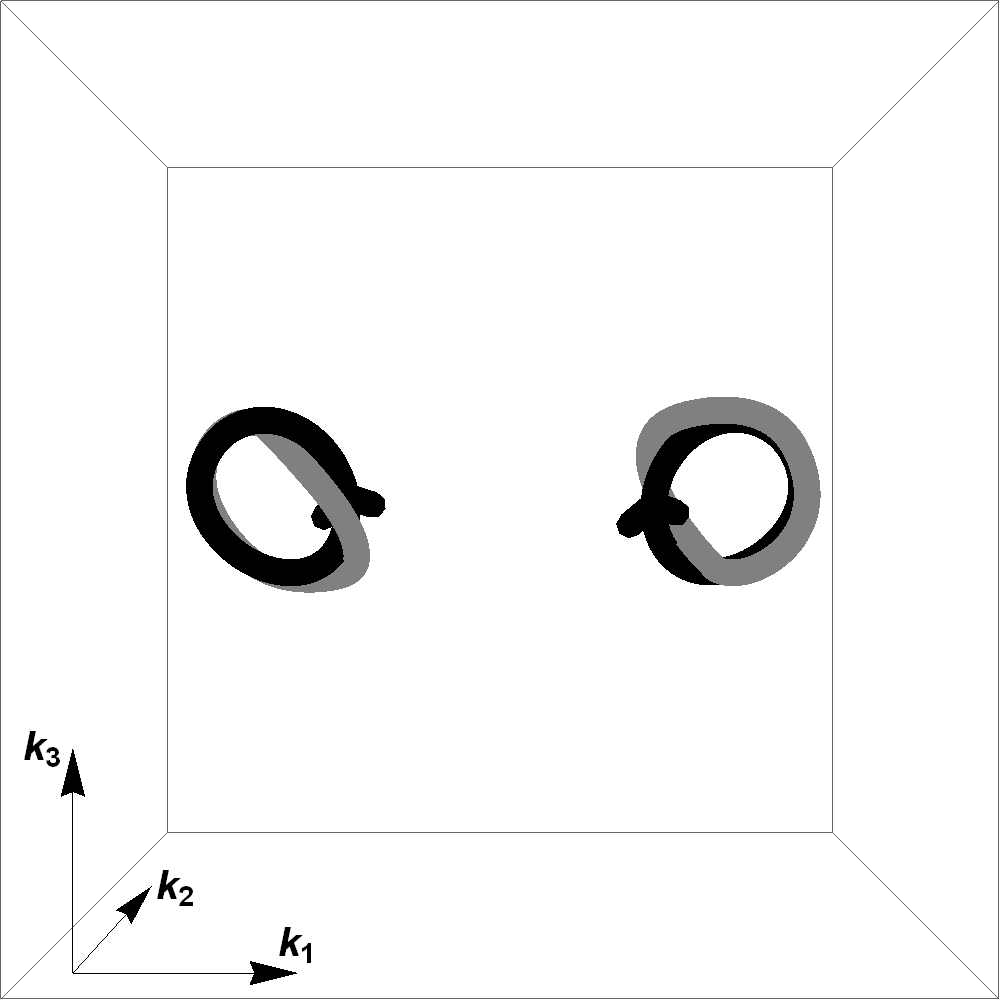}
\caption{Pontryagin manifold of the ground state map $\psi_{SC}$ associated to $H_{SC}$ in eq.~\eqref{eq:HSC}. The choice of regular value is $\psi_{SC}(3\pi/4,0,0)$.}\label{fig:hopfsc}
\end{figure}

The representative constructed in the previous paragraph is the ground state of Hamiltonians with terms connecting all sites of the real space lattice, albeit with amplitudes decreasing exponentially with distance. For the purpose of finding a tight-binding model that realizes the Hopf superconductor phase with only a finite hopping range, we use the freedom of homotopy to alter this representative. Thus, consider the following Bogoliubov-de Gennes Hamiltonian:
\begin{align}
H_\text{SC}(\mathbf{k})&:=\sum_{i=1}^3\big[\mathbf{z}^\dagger(\mathbf{k})\sigma_i\mathbf{z}(\mathbf{k})\big]\sigma_i,\label{eq:HSC}
\end{align}
where
\begin{align*}
\mathbf{z}(\mathbf{k})&:=\left(z_\uparrow(\mathbf{k}),z_\downarrow(\mathbf{k})\right)^T,\\
z_\uparrow(\mathbf{k})&:=\cos(k_1)+i\sin(k_2)\sin(k_1),\\
z_\downarrow(\mathbf{k})&:=\sin(k_3)\sin(k_1)\\
&+i(-\cos(2k_1)+\cos(k_2)+\cos(k_3)-5/2).
\end{align*}
This Hamiltonian resembles the ones introduced in \cite{hopf} and \cite{hopf2} for one half of the Brillouin zone ($k_1\le 0$) with its point-reflected partner on the other half ($k_1\ge0$).  An associated Pontryagin manifold of its ground state is shown in Fig.~\ref{fig:hopfsc}. Although the ground state map is not constant on the subtorus defined by $k_1=0$, the figure shows that the Chern number of its restriction vanishes (the preimage for $k_1=0$ is the empty set). Since two-dimensional symmetry class $C$ ground states are classified by their even Chern number -- often written as an entry $2\mathbb{Z}$ in the Periodic Table -- the restriction is equivariantly homotopic to the constant map.

\subsection{$\mathbb{Z}_2$ invariant}

In this section we introduce a general procedure of determining the $\mathbb{Z}_2$-invariant of a three-dimensional class $C$ superconductor with two (quasi-)bands. Given any (differentiable) Hamiltonian in this setting, this will provide a way to conclude whether or not the associated ground state $\psi$ is in the Hopf superconductor phase. Due to the constraint $\psi(-\mathbf{k})=\psi(\mathbf{k})$, the associated Jacobi matrix satisfies
\begin{align}
D\psi(-\mathbf{k})=-D\psi(\mathbf{k}).\label{eq:jacobi}
\end{align}
Three important properties of the preimage $\psi^{-1}(y)$ (where $y$ is regular as usual) can be deduced from this:
\begin{itemize}
\item[1.] Points with $\mathbf{k}=-\mathbf{k}$ are never contained in $\psi^{-1}(y)$.
\item[2.] Two points $\pm\mathbf{k}$ with $\mathbf{k}\ne-\mathbf{k}$ are either both contained in $\psi^{-1}(y)$ or none of them is.
\item[3.] If $\pm\mathbf{k}\in\psi^{-1}(y)$, then $\mathbf{k}$ and $-\mathbf{k}$ lie in different connected components.
\end{itemize}
The first statement follows from eq.~\eqref{eq:jacobi} together with the assumption that $y$ is regular. The second statement is an immediate consequence of the equivariance condition $\psi(\mathbf{k})=\psi(-\mathbf{k})$. For the third statement, let $\pm\mathbf{k}\in\psi^{-1}(y)$ and thus $\mathbf{k}\ne-\mathbf{k}$. If both points $\pm\mathbf{k}$ were to lie in the same connected component of $\psi^{-1}(y)$, then there would be a connecting path $\gamma:[-1,1]\to\psi^{-1}(y)$ with $\gamma(\pm1)=\pm\mathbf{k}$ and tangent vectors $\gamma'(t)$ that do not vanish for any $t$. Furthermore, $\gamma'(-1)=-c\gamma'(1)$ for some constant $c>0$ due to eq.~\eqref{eq:jacobi}. The framing of $\psi^{-1}(y)$ along $\gamma$ is given by two linearly independent vectors $u_1(t),u_2(t)$ orthogonal to $\gamma'(t)$ which obey the relation $u_i(-1)=-u_i(1)$ again due to eq.~\eqref{eq:jacobi}. The triplet of vectors $\{u_1(t),u_2(t),\gamma'(t)\}$ therefore has opposite orientations at $t=\pm1$, so there has to be some $t_0\in[-1,1]$ where the three vectors become linearly dependent, contradicting our assumptions. Thus, the two points $\pm\mathbf{k}$ have to lie on different connected components.

\begin{figure}[h]
\centering
\subfloat[]{\label{hopfsc-1}\includegraphics[width=0.2\textwidth]{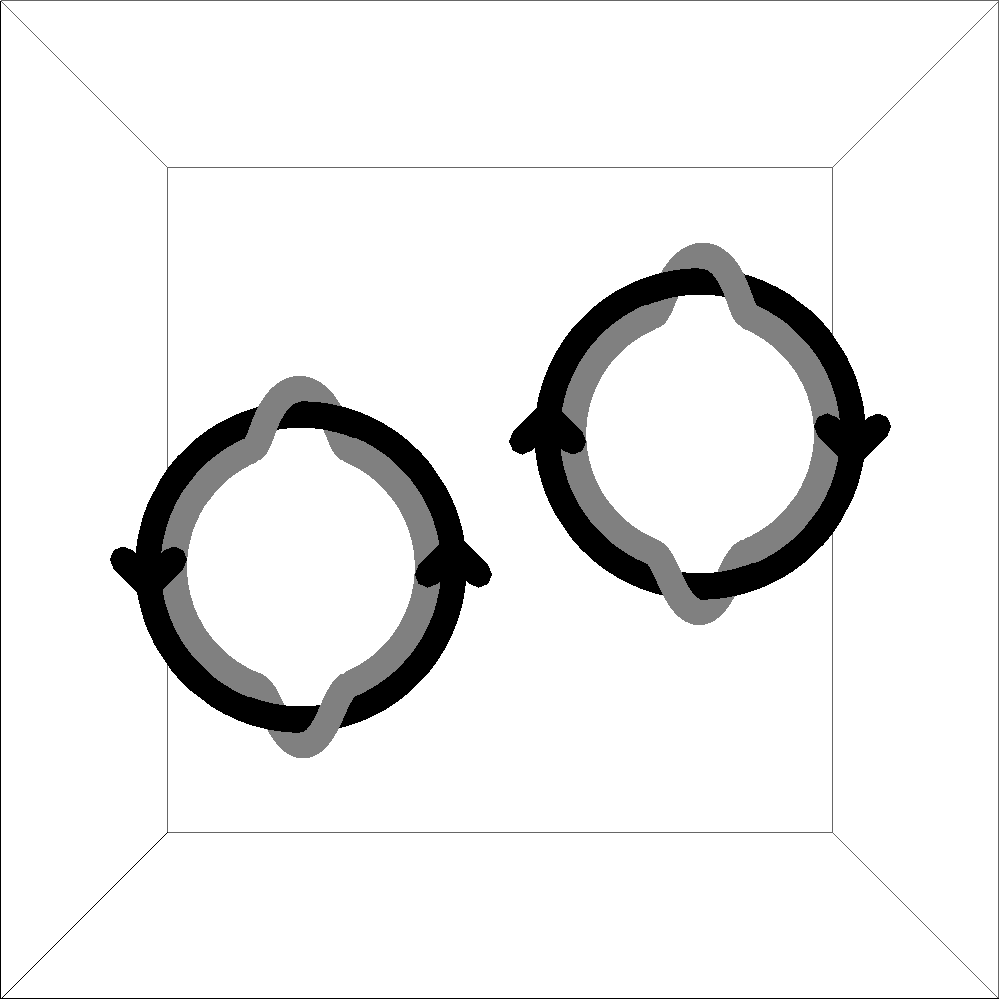}}\quad
\subfloat[]{\label{hopfsc-2}\includegraphics[width=0.2\textwidth]{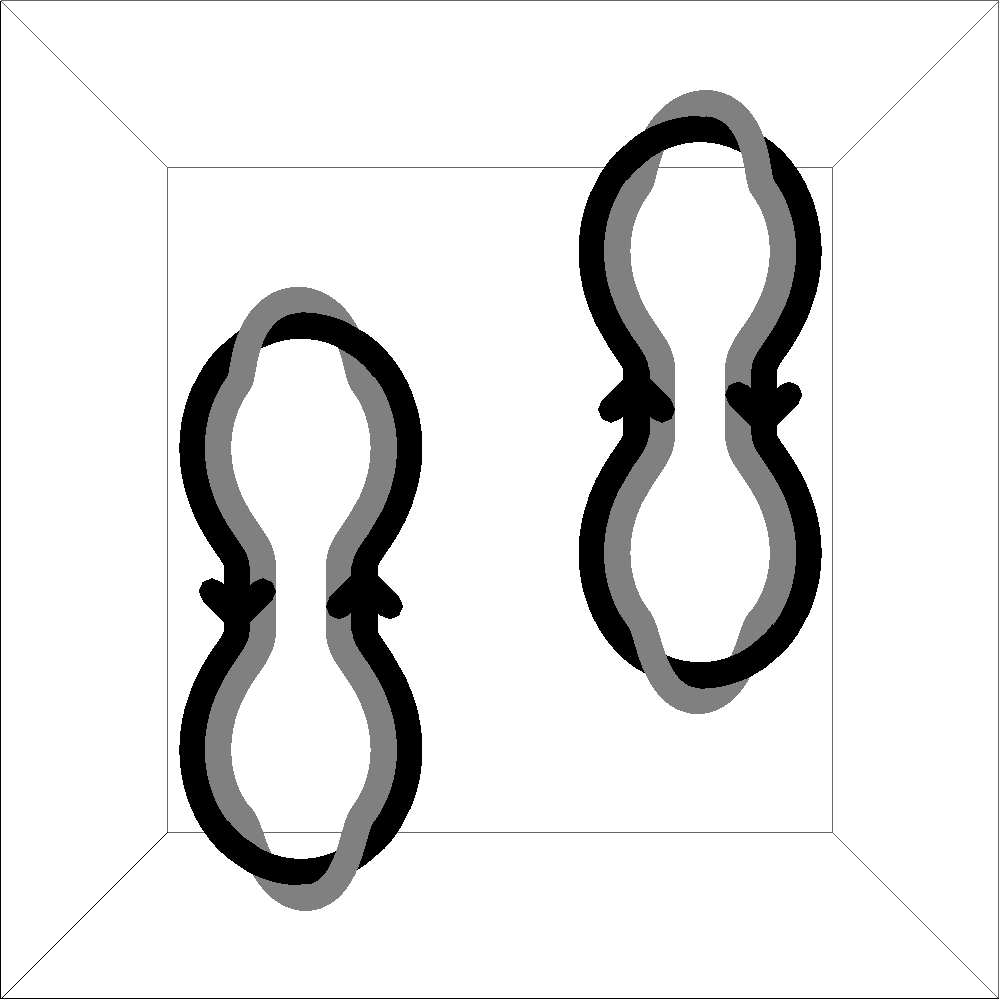}}\\
\subfloat[]{\label{hopfsc-3}\includegraphics[width=0.2\textwidth]{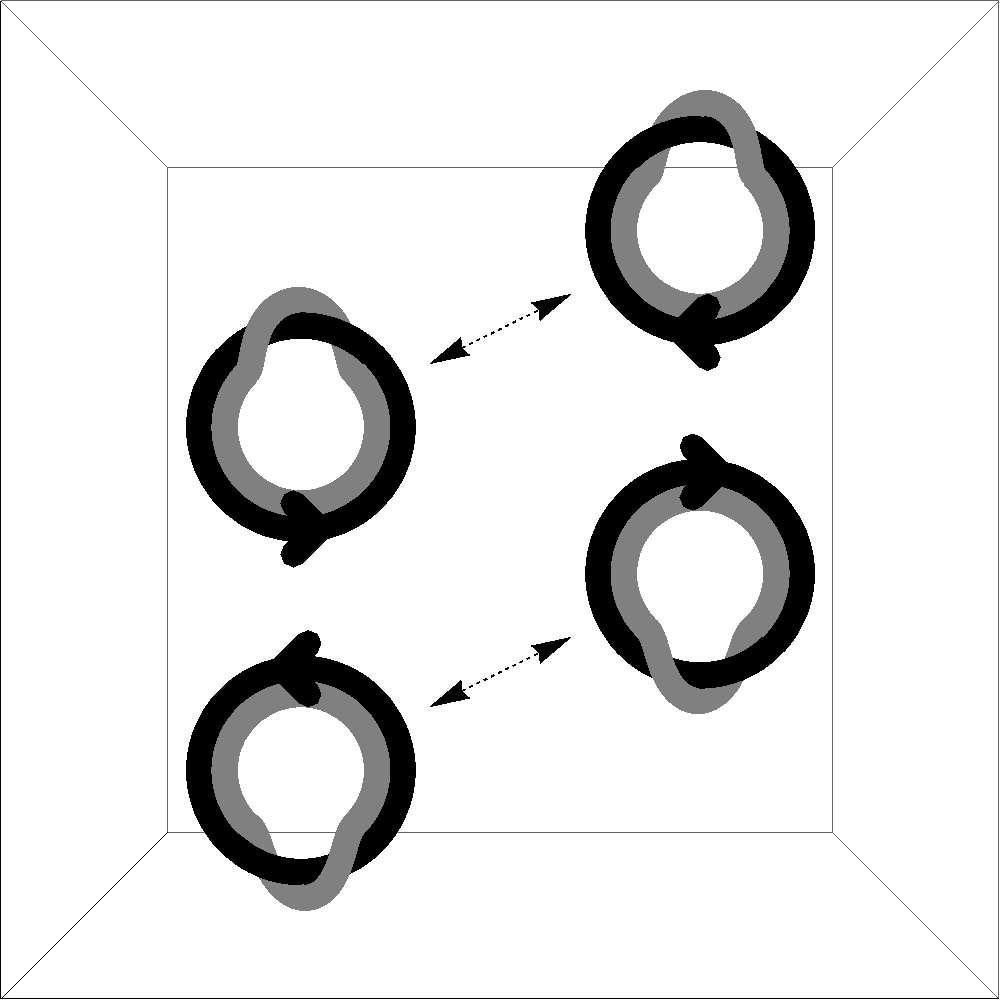}}
\caption[]{Triviality of superconductor with frame winding $\pm2$. \subref{hopfsc-1} Initial Pontryagin manifold with $m=1$ pair of connected components and frame winding $\nu_1=2$. \subref{hopfsc-2} Deformation leading to the pinching off in \subref{hopfsc-3} using the cobordism of Fig,~\ref{fig:cob2}. The resulting two pairs can be annihilated as indicated by the dashed arrows, avoiding the point $\mathbf{k}=0$ in the center.}\label{fig:hopfsc2}
\end{figure}

In other words, all connected components of $\psi^{-1}(y)$ come in pairs. This fact provides a definition of the $\mathbb{Z}_2$-invariant in the present context: Let the pairs of connected components be indexed by $i=1,\dots,m$ and let the frame winding number of one of the components in pair $i$ be $\nu_i$. Then the invariant $\nu\in\mathbb{Z}_2$ is given by
\begin{align}
\nu=\sum_{i=1}^m \nu_i\mod 2.
\end{align}
In particular, the ground state associated to the Hamiltonian in eq.~\eqref{eq:HSC} has $\mathbb{Z}_2$-invariant $\nu=1$, since $y$ can be chosen such that $m=1$ and $\nu_1=1$, see Fig.~\ref{fig:hopfsc}. Note that in the non-equivariant setting of symmetry class $A$, the two connected components with opposite frame winding could annihilate: First, form a single connected component using the cobordism in Fig.~\ref{fig:cob2} which then vanishes by the cobordism of  Fig.~\ref{fig:cob1}. However, in symmetry class $C$ this annihilation would have to occur at a point with $\mathbf{k}=-\mathbf{k}$, which cannot happen because these points are excluded from preimages of regular values (also, only even numbers of connected components are permitted). On the other hand, if $m=1$ with $\nu_1=2$, the $\mathbb{Z}_2$-invariant is $\nu=0$ and thus trivial. This is illustrated in Fig.~\ref{fig:hopfsc2}: The pair of connected components with frame winding $\pm2$ can be split via an equivariant framed cobordism into two pairs (so four in total) with frame windings $\pm1$ respectively. These can then mutually annihilate away from points with $\mathbf{k}=-\mathbf{k}$. 

Using these two insights, the general nature of the $\mathbb{Z}_2$-invariant becomes evident: The $m$ pairs of connected components with frame windings $\nu_1,\dots,\nu_m$ of a general Pontryagin manifold can be merged into a single pair with windings $\pm\sum_i\nu_i$ by repeatedly applying the cobordism of Fig.~\ref{fig:cob2}. If $\sum_i\nu_i=0$, then the two constituents may be annihilated individually. Otherwise, if $\sum_i\nu_i$ is even, then a pair with windings $\pm2$ can be split off and annihilated according to the procedure outlined previously (and illustrated in Fig.~\ref{fig:hopfsc2}). Iterating this process, a framed cobordism to the empty set is found, thereby establishing a deformation to the trivial phase (containing the constant ground state map of the atomic limit). On the other hand, if $\sum_i\nu_i$ is odd, the process terminates at a single pair of components with windings $\pm1$, which cannot be reduced any further.

\subsection{Boundary of a Hopf superconductor}

\begin{figure}[h]
\centering
\subfloat[]{\label{hopfsc-x}\includegraphics[width=0.2\textwidth]{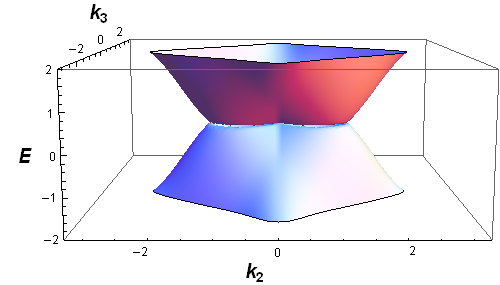}}\quad
\subfloat[]{\label{hopfsc-y}\includegraphics[width=0.2\textwidth]{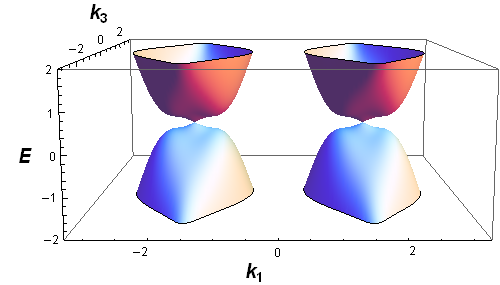}}\\
\subfloat[]{\label{hopfsc-z}\includegraphics[width=0.2\textwidth]{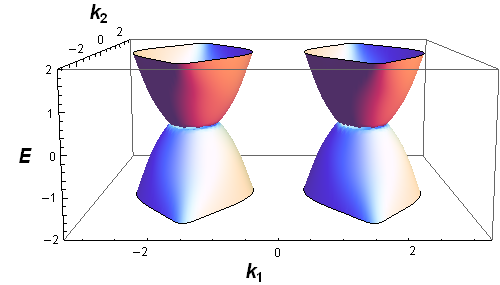}}\quad
\caption[]{Energy spectrum of the Hopf superconductor Hamiltonian in eq.~\eqref{eq:HSC} in a $16$-site thick slab geometry with boundary orthogonal to \subref{hopfsc-x} the $(1,0,0)$ direction, \subref{hopfsc-x} the $(0,1,0)$ direction and \subref{hopfsc-x} the $(0,0,1)$ direction. Shown are the two energy levels closest to the Fermi energy $E=0$. The gap closes in all three cases, creating the following Fermi surfaces of zero modes: \subref{hopfsc-x} Figure eight pattern, \subref{hopfsc-y} two Dirac points and \subref{hopfsc-z} two rings.}\label{fig:hopfsc-boundary}
\end{figure}

We use the same procedure of introducing boundaries as for the Hopf-Chern insulators from the previous part of this paper. As to be expected from the principle of bulk-boundary correspondence, the spectrum with boundary is gapless (see Fig.~\ref{fig:hopfsc-boundary}). Apart from some modifications to ensure equivariance, the main construction step of the Hopf superconductor Hamiltonian in eq.~\eqref{eq:HSC} is a doubling of the Hopf insulator models used in \cite{hopf} and \cite{hopf2}. Since we chose the $k_1$-direction for this purpose, the spectrum with boundary transverse to this direction resembles two copies of the one obtained in \cite{hopf}: For a boundary orthogonal to $(0,0,1)$, there are two rings of zero modes (Fig.~\ref{fig:hopfsc-boundary}\subref{hopfsc-z}) while for one orthogonal to $(0,1,0)$ there are two Dirac cones  (Fig.~\ref{fig:hopfsc-boundary}\subref{hopfsc-y}. With a boundary orthogonal to the $(1,0,0)$-direction however, a novel spectrum is obtained where the gap closes along two intersecting rings, see Fig.~\ref{fig:hopfsc-boundary}\subref{hopfsc-x}.

The features of the Fermi surfaces change for other representatives in the Hopf superconductor phase obtained by applying homotopies to $H_{SC}$ that leave the bulk gap open, but the fact that the gap closes when a boundary is present seems to be robust. A general proof of bulk-boundary correspondence, for which a lot of progress has been made in the stable regime \cite{hatsugai,kellendonk,gurarie,graf}, would be desirable in this context. An approach that is valid outside the stable regime has been given in \cite{mong}, but only for nearest-neighbor hopping Hamiltonians, excluding the Hopf superconductor and all Hopf-Chern insulators.

\section{Discussion}

The restriction to two-band models required for the topological phases discussed in this paper deserves justification. For the Hopf superconductor, being described from the start by an effective theory (BCS theory), the consequences presented here (gapless surface modes) persist as long as this effective description is valid. In a similar vein, the two-band description leading to Hopf-Chern topological insulators and the accompanying phenomenology is only valid if there is an energy gap to any further valence or conduction bands such that there is no significant hybridization. The precise meaning of ``significant'' is a subject of further research.

Another point worth discussing is the robustness of the presented topological phases to disorder. First numerical results indicate that the closing of the energy gap is indeed a robust feature in the presence of translation-invariance breaking disorder, in agreement with the findings in \cite{hopf2}. However, as emphasized in \cite{hopf2}, the nature of the zero modes may change from extended to localized after introducing disorder.

\section{Conclusion}

We introduced novel topological phases of gapped, translation invariant, two-band free fermion systems in symmetry classes $A$ and $C$. These phases occur in three spatial dimensions and they lie outside the scope of the Periodic Table of topological insulators and superconductors. In class $A$, these are the infinitely many Hopf-Chern topological insulator phases which are truly 3D phases that can be thought of as a combination of layered Chern insulators and Hopf insulators. In class $C$, there is only one non-trivial topological phase, the Hopf superconductor, which is a superconducting cousin of the Hopf insulator. We have introduced geometrical diagnostics for all of these novel phases and for a large subset of them we were able to provide simple tight-binding Hamiltonians, whose energy spectra we investigated in the presence of a boundary. In all cases we found these spectra to be gapless, hinting at a form of bulk-boundary correspondence valid beyond the stable regime.

\subsection{Acknowledgments}

The author would like to acknowledge financial support through DFG grant ZI 513/2-1. Furthermore, the author is grateful for helpful discussions with Dominik Ostermayr, who pointed out the simplified derivation of the Hopf superconductor $\mathbb{Z}_2$-classification presented here, and Martin Zirnbauer.

\appendix

\section{Appendix: Framed cobordisms}

\begin{figure}[h!]
\centering
\begin{tikzpicture}
    \def\w{0.2\textwidth}
    \def\s{4mm}
    \node[anchor=east] at (0,0)
    {\includegraphics[width=\w]{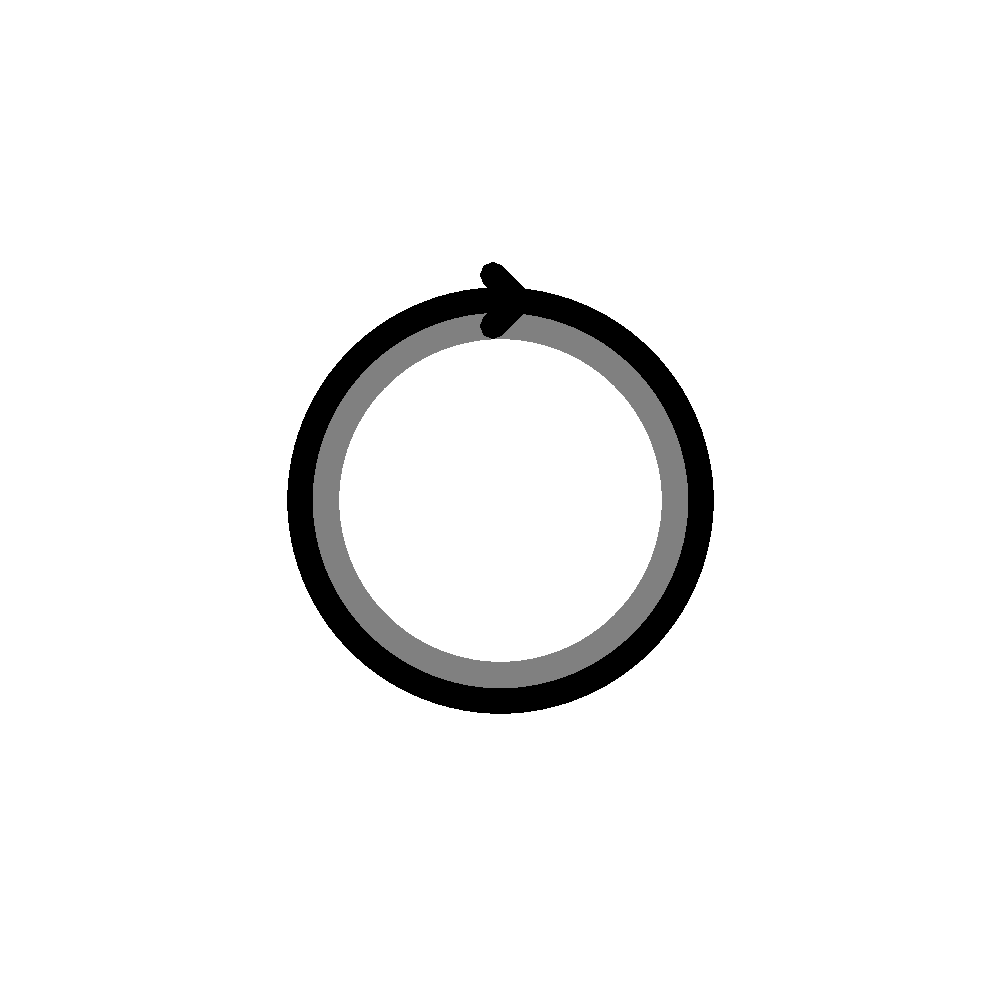}};
    \node[anchor=west] at (0,0)
    {\includegraphics[width=\w]{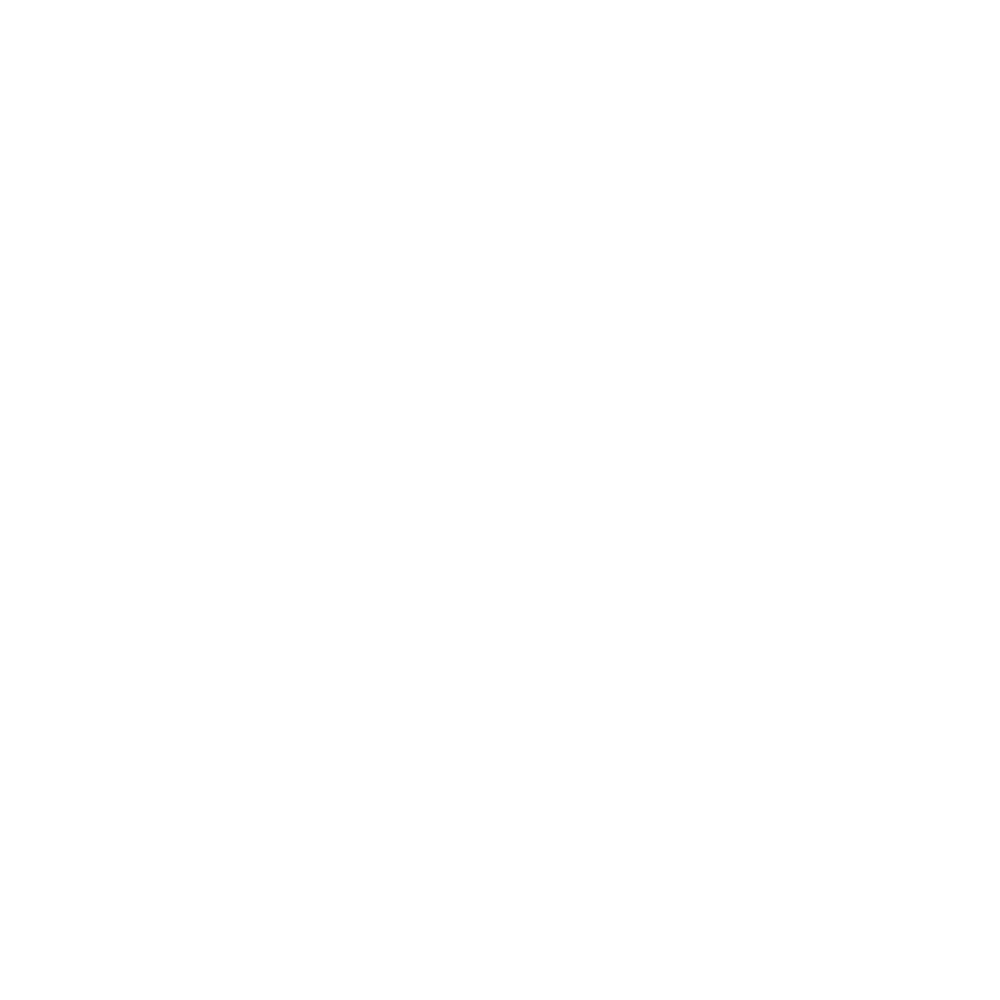}};
    \node at (0,0)
    {$\sim$};
\end{tikzpicture}
\caption{Contractible submanifolds with no frame windings are cobordant to the empty set.}\label{fig:cob1}
\end{figure}

\begin{figure}[h!]
\centering
\begin{tikzpicture}
    \def\w{0.2\textwidth}
    \def\s{4mm}
    \node[anchor=east] at (0,0)
    {\includegraphics[width=\w]{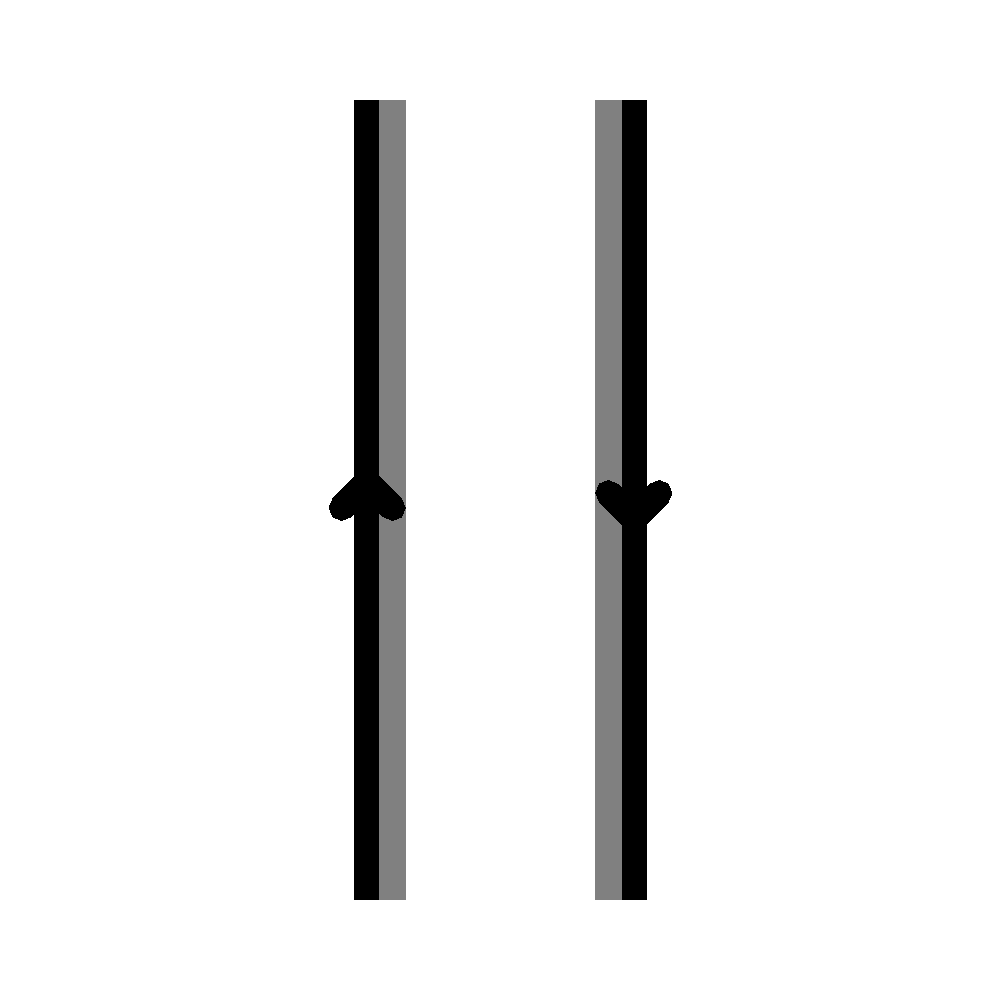}};
    \node[anchor=west] at (0,0)
    {\includegraphics[width=\w]{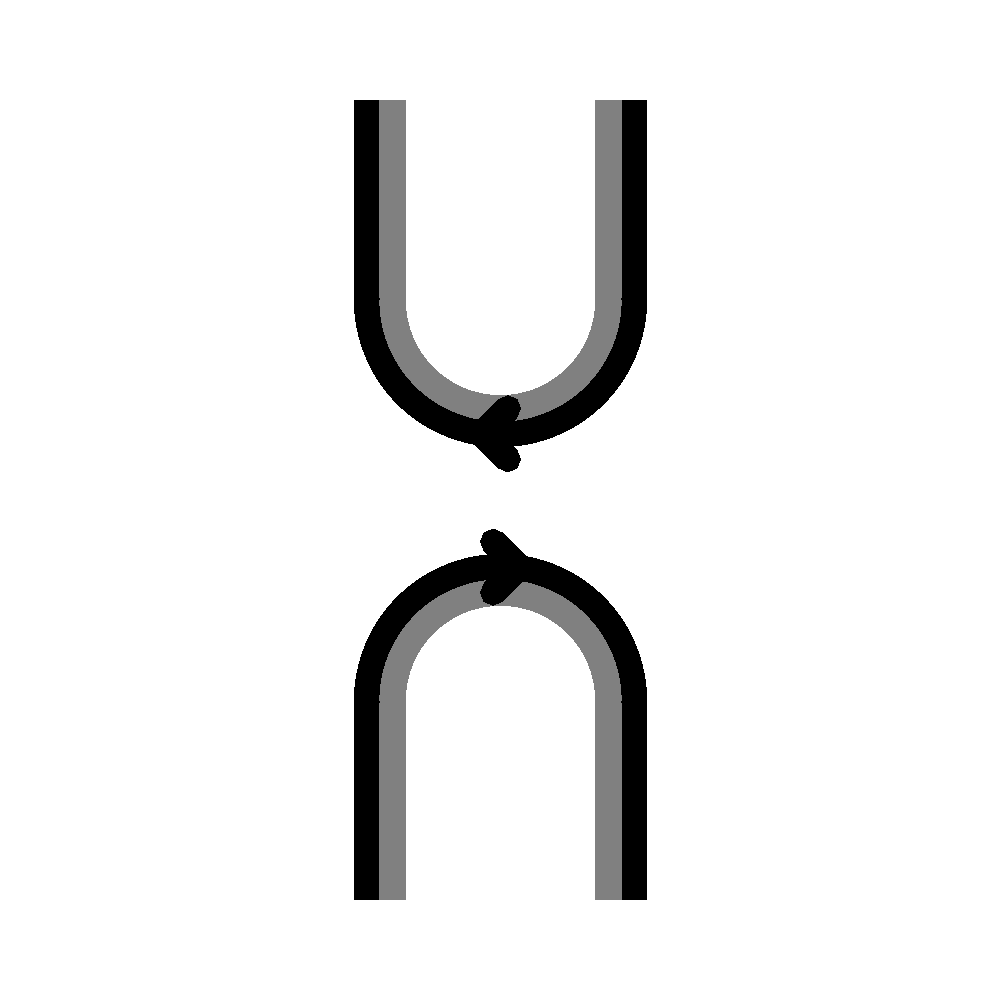}};
    \node at (0,0)
    {$\sim$};
\end{tikzpicture}
\caption{Saddle cobordism.}\label{fig:cob2}
\end{figure}

\begin{figure}[h!]
\centering
\begin{tikzpicture}
    \def\w{0.2\textwidth}
    \def\s{4mm}
    \node[anchor=east] at (0,0)
    {\includegraphics[width=\w]{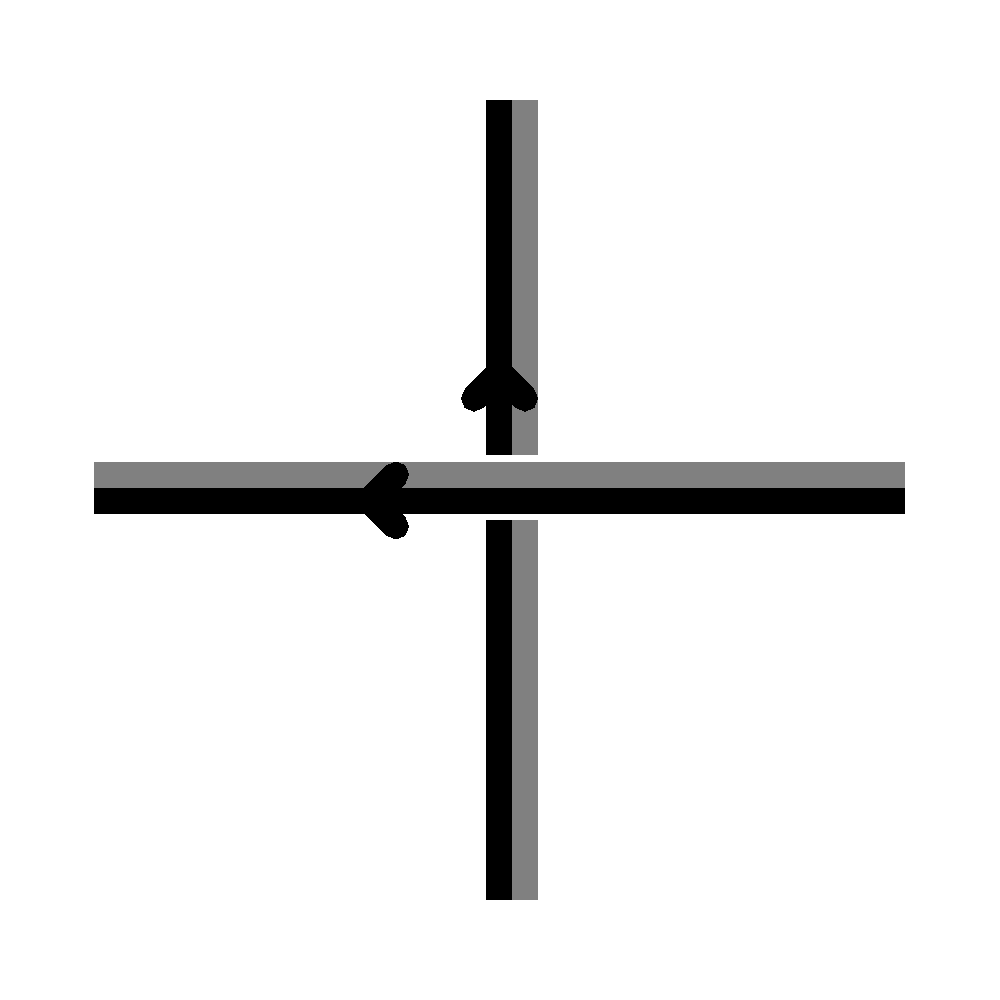}};
    \node[anchor=west] at (0,0)
    {\includegraphics[width=\w]{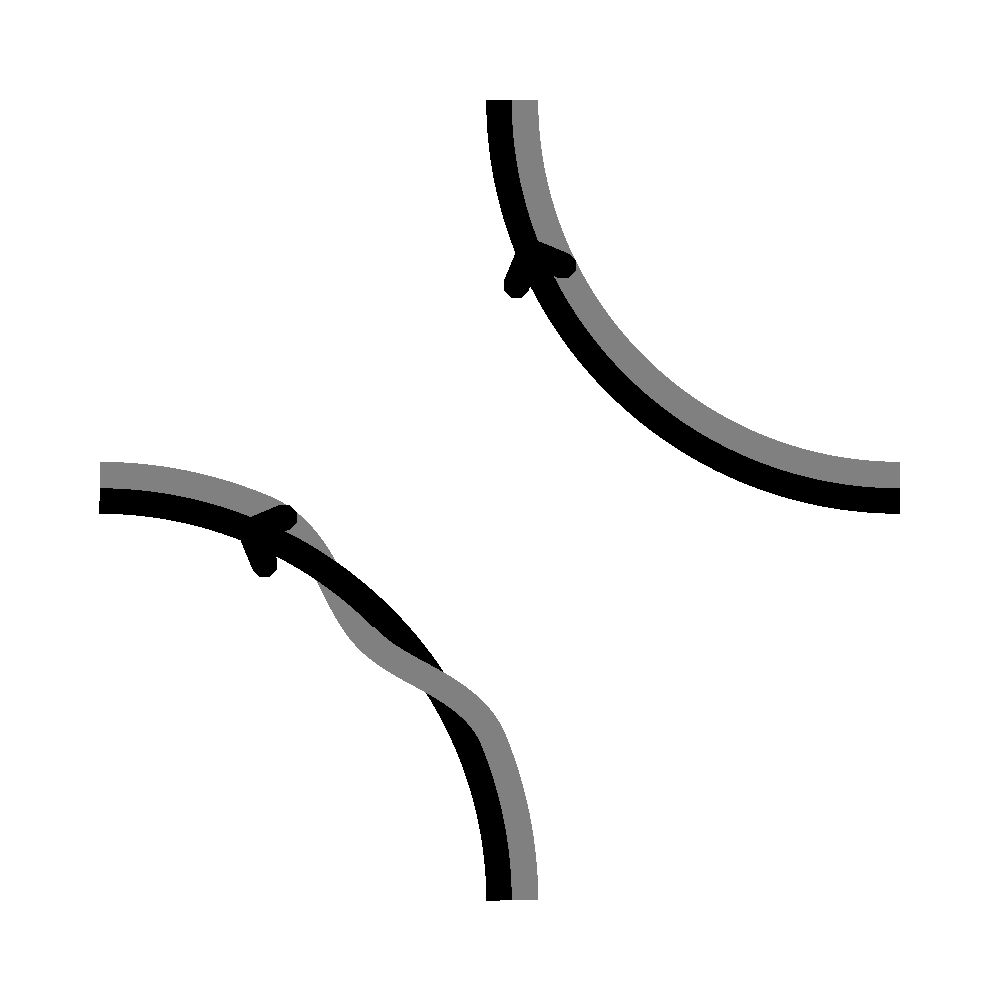}};
    \node at (0,0)
    {$\sim$};
\end{tikzpicture}
\caption{Crossings can be undone via a cobordism at the expense of introducing a frame winding.}\label{fig:cob3}
\end{figure}

In this part, we illustrate some important examples of framed cobordisms in the setting of one-dimensional framed submanifolds. This setting is relevant for the case of maps $T^3\to\text{Gr}_1(\mathbb{C}^2)\simeq S^2$ describing symmetry class~$A$ insulators in three spatial dimensions.

First and foremost, all continuous deformations of the submanifold or its framing are cobordisms. However, there are more possibilities, as shown in Figs.~\ref{fig:cob1}, \ref{fig:cob2} and \ref{fig:cob3}. For illustrations of the cobordisms with boundaries as displayed in these figures, we refer to \cite{triple}.


\begin{thebibliography}{10}
\bibitem{bernevig} B.~A.~Bernevig, T.~A.~Hughes, S.~C.~Zhang, \textit{Quantum Spin Hall Effect and Topological Phase Transition in HgTe Quantum Wells}, Science~\textbf{314},~1757~(2006)
\bibitem{fukane} L.~Fu, C.~L.~Kane, \textit{Topological insulators with inversion symmetry}, Phys.~Rev.~B~\textbf{76},~045302~(2007)
\bibitem{zhang} H.~Zhang, C.~X.~Liu, X.~L.~Qi, X.~Dai, Z.~Fang, S.~C.~Zhang, \textit{Topological insulators in Bi2Se3, Bi2Te3 and Sb2Te3 with a single Dirac cone on the surface}, Nat.~Phys.~\textbf{5},~438~(2009)
\bibitem{haldane} F.~D.~M.~Haldane, \textit{Model for a Quantum Hall Effect without Landau Levels: Condensed-Matter Realization of the "Parity Anomaly"}, Phys. Rev. Lett.~\textbf{61}, 2015 (1988)
\bibitem{molenkamp} M.~K\"onig, S.~Wiedmann, C.~Br\"une, A.~Roth, H.~Buhmann, L.~W.~Molenkamp, X.~L.~Qi, S.~C.~Zhang, \textit{Quantum Spin Hall Insulator State in HgTe Quantum Wells}, Science~\textbf{318},~766~(2007)
\bibitem{hasan} D.~Hsieh, D.~Qian, L.~Wray, Y.~Xia, Y.~S.~Hor, R.~J.~Cava, M.~Z.~Hasan, \textit{A topological Dirac insulator in a quantum spin Hall phase} Nature~\textbf{452},~970~(2008)
\bibitem{hasan2} Y.~Xia, D.~Qian, D.~Hsieh, L.~Wray, A.~Pal, H.~Lin, A.~Bansil,
D.~Grauer, Y.~S.~Hor, R.~J.~Cava, M.~Z.~Hasan, \textit{Observation of a large-gap topological-insulator class with a single Dirac cone on the surface}, Nat.~Phys.~\textbf{5},~398~(2009)
\bibitem{chang} C.-Z.~Chang et al \textit{Experimental Observation of the Quantum Anomalous Hall Effect in a Magnetic Topological Insulator}, Science~\textbf{340}, 6129 (2013) 
\bibitem{kitaev} A.~Kitaev, \textit{Periodic table for topological insulators and superconductors}, AIP Conf. Proc. 1134, 22 (2009)
\bibitem{kz} R.~Kennedy, M.~R.~Zirnbauer, \textit{Bott Periodicity for $\mathbb{Z}_2$ symmetric ground states of gapped free-fermion systems}, Communications in Mathematical Physics~\textbf{342}, 909-963 (2016)
\bibitem{hhz} P.~Heinzner, A.~Huckleberry, M.~R.~Zirnbauer, \textit{Symmetry classes of disordered fermions}, Commun.~Math.~Phys.~\textbf{257},~725~(2005)
\bibitem{hopf} J.~E.~Moore, Y.~ Ran, X.-G.~Wen, \textit{Topological Surface States in Three-Dimensional Magnetic Insulators}, Phys.~Rev.~Lett.~\textbf{101}, 186805 (2008)
\bibitem{hopf2} D.-L.~Deng, S.-T.~Wang, C.~Shen, and L.-M.~Duan, \textit{Hopf insulators and their topologically protected surface states}, Phys.~Rev.~B~\textbf{88}, 201105(R) (2013)
\bibitem{charlie} R.~Kennedy, C.~Guggenheim, \textit{Homotopy theory of strong and weak topological insulators}, Phys. Rev.~B~\textbf{91}, 245148 (2015)
\bibitem{pontryagin} L.~S.~Pontryagin, \textit{A classification of mappings of the three-dimensional complex into the two-dimensional sphere}, Mat. Sbornik (Recueil Mathematique N. S.) \textbf{9}, 331 (1941)
\bibitem{botttu} R.~Bott, L.~W.~Tu, \textit{Differential Forms in Algebraic Topology}, Springer, New York (1982)
\bibitem{triple} D.~DeTurck, H.~Gluck, R.~Komendarczyk, P.~Melvin, C.~Shonkwiler and D.~S.~Vela-Vick \textit{Triple linking numbers, ambiguous Hopf invariants and integral formulas for three-component links}, Mat. comtemp.~\textbf{34}, 251-283 (2008). 
\bibitem{milnor} J.~Milnor, \textit{Topology from the Differentiable Viewpoint}, University Press of Virginia, Charlottesville~(1965)
\bibitem{hatsugai} Y.~Hatsugai, \textit{Chern number and edge states in the integer quantum Hall effect}, Phys.~Rev.~Lett.~\textbf{71}, 3697 (1993)
\bibitem{kellendonk} J.~Kellendonk, T.~Richter, H.~Schulz-Baldes, \textit{Edge current channels and Chern numbers in the integer quantum Hall effect} Rev.~Math.~Phys.~\textbf{14}, 87 (2002)
\bibitem{gurarie} A.~M.~Essin, V.~Gurarie, \textit{ Bulk-boundary correspondence of topological insulators from their Green's functions}, Phys.~Rev.~B \textbf{84}, 125132 (2011)
\bibitem{graf} G.~M.~Graf, M.~Porta, \textit{Bulk-Edge Correspondence for Two-Dimensional Topological Insulators}, Communications in Mathematical Physics, \bibitem{mong} R. S. K. Mong, V. Shivamoggi, \textit{Edge states and the bulk-boundary correspondence in Dirac Hamiltonians} Phys.~Rev.~B~\textbf{83}, 125109 (2011)
\bibitem{hatcher} A.~Hatcher, \textit{Algebraic Topology}, Cambridge~University~Press, Cambridge~(2002)
\bibitem{tft} X.-L.~ Qi, T.~ L.~Hughes, S.-C.~Zhang, \textit{Topological field theory of time-reversal invariant insulators}, Phys. Rev. B~\textbf{78}, 195424 (2008)
\end{thebibliography}
\end{document}